\documentclass[fleqn,10pt]{wlscirep}
\usepackage[utf8]{inputenc}
\usepackage[T1]{fontenc}
\usepackage{braket}
\usepackage{amsmath}
\usepackage{amssymb}
\usepackage[resetlabels]{multibib}
\newcites{supp}{Supplementary References}

\title{Valley enhanced Rabi frequency in n-type planar Silicon-MOS quantum dot }
\author[1,2,*]{Xunyao Luo}
\author[1,2]{Xander Peetroons}
\author[2]{Tsung-Yeh Yang}
\author[2]{Ruben M. Otxoa}
\author[2]{Normann Mertig}
\author[3]{Sofie Beyne}
\author[3]{Julien Jussot}
\author[3]{Yosuke Shimura}
\author[3]{Clement Godfrin}
\author[3]{Bart Raes}
\author[3]{Roy Li}
\author[3,5]{Roger Loo}

\author[3]{Sylvain Baudot}
\author[3]{Stefan Kubicek}
\author[3]{Shuchi Kaushik}
\author[3]{Danny Wan}
\author[3,6,7]{Kristiaan De Greve}
\author[4]{Takuma Kuno}
\author[4]{Takeru Utsugi}
\author[4]{Noriyuki Lee}
\author[4]{Itaru Yanagi}
\author[4]{Toshiyuki Mine}
\author[4]{Satoshi Muraoka}
\author[4]{Hideo Arimoto}
\author[4]{Shinichi Saito}
\author[4]{Digh Hisamoto}
\author[4]{Ryuta Tsuchiya}
\author[4]{Hiroyuki Mizuno}
\author[1,2]{Charles Smith}
\author[2]{Andrew Ramsay}
\affil[1]{Cavendish Laboratory, Department of Physics, University of Cambridge, Cambridge CB3 0US, United Kingdom}
\affil[2]{Hitachi Cambridge Laboratory, J. J. Thomson Avenue, Cambridge CB3 0US, United Kingdom}
\affil[3]{IMEC, Leuven, Belgium}
\affil[4]{Research and Development Group, Hitachi, Ltd., Kokubunji, Tokyo 185-8601, Japan}
\affil[5]{Dept. of Solid-State Sciences, Ghent University, Krijgslaan 285,  9000 Ghent, Belgium}
\affil[6]{Proximus chair in Quantum Science and Engineering, Department of Electrical Engineering (ESAT-MNS), KU Leuven, Leuven, Belgium}
\affil[7]{Dept. of electrical engineering (ESAT-MNS), KU Leuven, Leuven, Belgium}
\affil[*]{Corresponding author: Xunyao Luo (xl527@cam.ac.uk)}

\begin{abstract}
Electron spin resonance spectroscopy (ESR) of a single electron in planar Si-MOS quantum dot is reported in the vicinity of a valley level anti-crossing. A number of one and two-photon resonances are observed due to mixing of magnetic spin-flip and electric valley-flip transitions. This allows the reconstruction of the energy-level diagram of a four state system with two valley and two spin states. Near the anti-crossing, an enhancement of the Rabi frequency is observed. This is attributed to an electric-dipole transition activated by admixing of the upper energy level due to inter-valley spin coupling. The electric-dipole transition may be driven via capacitive coupling between the ESR antenna, and the confinement gate. To characterize spin-valley  coupling responsible for the enhancement, we  
measure the anisotropy of the g-factor difference between the two valley states, the mean g-factor and the inter-valley spin coupling for both in and out-of-plane magnetic fields.  The inter-valley spin coupling is strongly modulated by the direction of the B-field, and is strongest for out-of-plane B-field, consistent with an in-plane spin-valley field. 
In principle, this  strong Electric dipole spin resonance (EDSR) effect could be utilized for fast all-electrical spin control in small-scale devices. 

\end{abstract}
\begin{document}

\flushbottom
\maketitle

\thispagestyle{empty}

\section*{Introduction}

Electron spins in silicon quantum dots represent one of the most promising platforms for scalable quantum computing, combining long coherence times with compatibility with advanced CMOS fabrication processes\cite{RevModPhys.85.961,Chatterjee2021,RevModPhys.95.025003}. In bulk silicon, the spin-orbit interaction is quite weak, and the g-factor is close to two. Hence, direct control is limited to electron spin resonance (ESR), where an oscillating magnetic field drives spin rotations\cite{Steinacker2025}. However, since the magnetic coupling strength is intrinsically weak, the gate speed is limited.  %Enhancing spin rotation rates without compromising device scalability remains a central challenge.
Faster control can be achieved using a synthetic spin-orbit interaction using the magnetic-field gradient generated by a micromagnet at the expense of increased charge sensitivity \cite{Pioro-Ladrière2008,Klemt2023}. However, micromagnets require processes that are not standard in many foundries\cite{Koch2025}. Faster control can also be achieved using the exchange interaction at the expense of a more complex qubit encoding, and quite elaborate control sequences\cite{PhysRevApplied.22.044057,Weinstein2023}. 

Silicon has a multivalley conduction-band introducing an additional degree of freedom beyond spin. At an atomically sharp Si/SiO\textsubscript{2} interface, the bulk inversion symmetry is broken, and the two lowest-energy valleys aligned along the out-of-plane directions are split by the interface potential, giving rise to the valley splitting. In the presence of spin–orbit coupling, the valley and spin states can hybridize, especially when their energy separation becomes comparable to the Zeeman energy \cite{Yang2013}. Such spin–valley hybridization can significantly modify the energy relaxation rate\cite{Yang2013,GuoPhysRevLett.124.257701}, and result in enhancements of the gate driven Rabi frequency\cite{Klemt2023,EDSRLeti}.

Previous studies have explored this mechanism mainly in corner-dots \cite{EDSRLeti,Klemt2023} or in asymmetric confinement geometries \cite{Jock2018}, where strong interface-induced spin–orbit coupling or micromagnet-induced field gradients are deliberately introduced to enable electrically driven spin resonance (EDSR) \cite{EDSRLeti}, or in Si/SiGe quantum dots where small valley-splittings are common \cite{Cai2023}. In contrast, planar MOS quantum dots, are expected to exhibit larger valley splittings due to the large potential step at the Si/SiO$_2$ interface, and the triangular vertical potential well. Consequently, they  are expected to require a micromagnet to perform EDSR \cite{Ma_PhysRevApplied.21.014044}. However, recently small valley-splittings and EDSR resonances without micromagnet have been reported in planar Si-MOS devices similar to the device used here\cite{tomić2025longcoherencesiliconspin}.

In this work, we demonstrate that at magnetic fields close to a valley anti-crossing a strong enhancement of the Rabi frequency is observed for electron spin-resonance driven via an antenna. This is interpreted as an EDSR contribution to the Rabi frequency activated by admixing due to inter-valley spin coupling. The device has no micromagnet, and is a planar Si-MOS quantum dot with poly-Si gates for low strain. Nonetheless it appears that a significant EDSR effect is observed.

In addition, detailed spectroscopy of the spin-valley interactions in the vicinity of the anti-crossing are reported. This includes the valley dependent anisotropy of the g-factor, the anisotropy of the inter-valley spin coupling, and spin relaxation measurements.  This work presents evidence that the variability in the g-factor is dominated by  spin-valley physics. 

\section*{Results}

\subsection*{Device specification and spin-to-charge conversion readout}

We use a single-electron spin qubit in a natural-silicon MOS planar double quantum dot device, fabricated on a 300-mm silicon-on-insulator (SOI) wafer at IMEC\cite{peetroons2025highfidelityqubitcontrol,Elsayed2024,LiIMEC}. The device consists of undoped planar silicon on top of a buried oxide, with quantum dot–defining gate layers patterned above, as illustrated schematically in Fig.~\ref{fig:1}(a)\cite{Elsayed2024}. The device consists of an RF-SET charge sensor, and a double quantum dot, operated as a single quantum dot. We use Elzerman energy-selective spin readout\cite{Elzerman2004}. This discriminates between the high (the three excited states) and low (the ground) energy-levels at the readout point of the experiment. The electron temperature of Te=163mK is relatively high resulting in imperfect initialization and read-out\cite{Keith_2019}, which helps to populate the first two excited state. Elzerman readout in the presence of valley states is discussed in more detail in the Supplementary information.

Coherent control of the electron spin is achieved using electron spin resonance (ESR). Microwave (MW) pulses are applied through an aluminium antenna placed in proximity to the quantum dot. The microwave control signals are generated using an IQ mixer, specifically Quantum Machines Octave, where an intermediate frequency $f_{IF}\leq350\,$MHz, is mixed with a local oscillator (LO) frequency $f_{LO}\approx 16.0\sim17.5\,\mathrm{GHz}$ to produce three principal output tones: $f_{LO}$, $f_{LO}+f_{IF}$, and $f_{LO}-f_{IF}$. The mixer in the Octave is calibrated at each IF-frequency, so that both the LO frequency $f_{LO}$ and the lower sideband $f_{LO}-f_{IF}$ are suppressed by  $-40$ dB, compared to the main microwave tone at $f_{MW}=f_{LO}+f_{IF}$.

\subsection*{Anticrossing in ESR spectra due to inter-valley spin-orbit coupling}

\begin{figure}[h!]
\centering
\includegraphics[width=\linewidth]{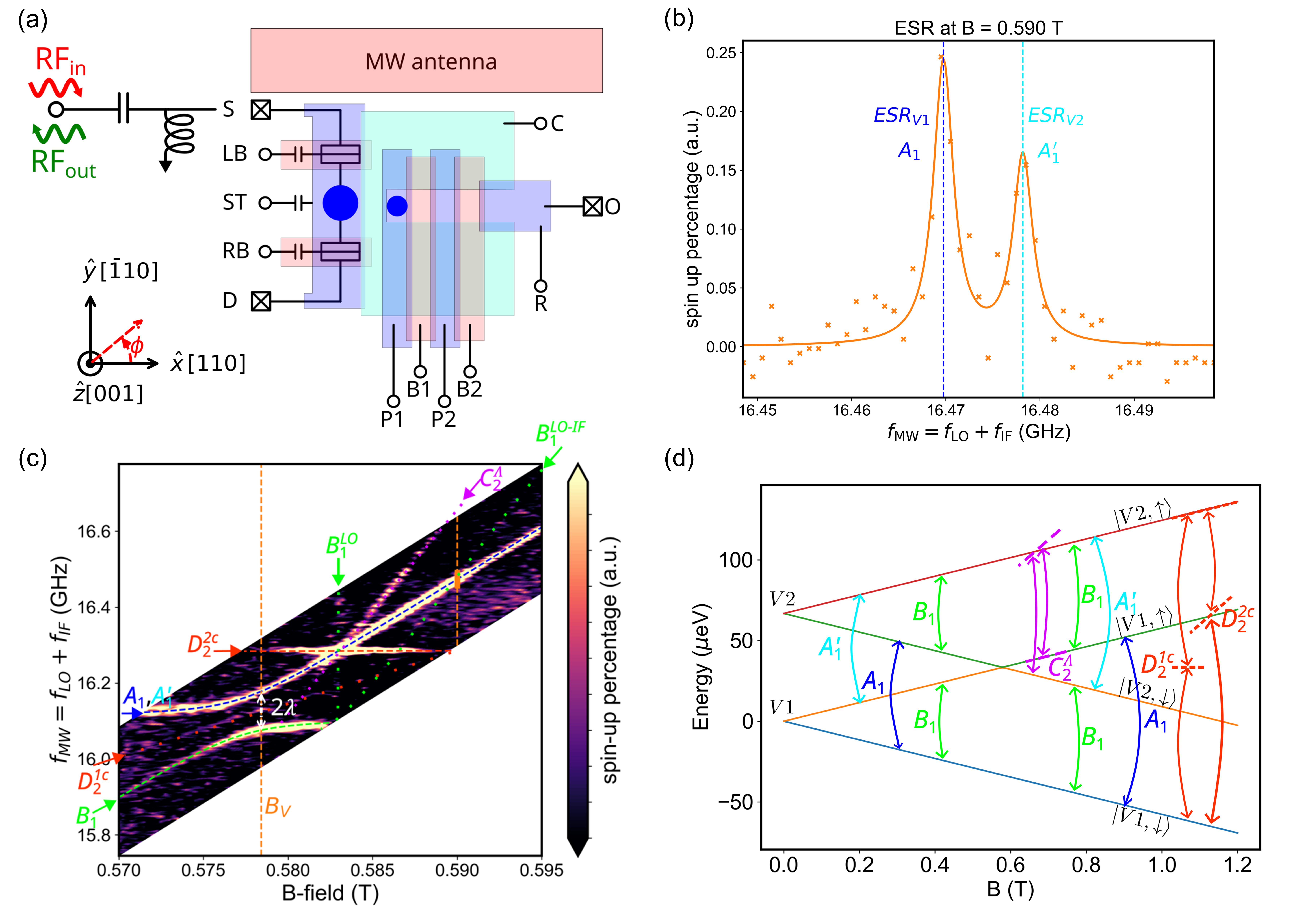}
\caption{Device and spin--valley spectroscopy. (a) Device schematic illustrating electron spin resonance (ESR) excitation using microwave (MW) pulses applied through an aluminum antenna, with spin readout performed via the Elzerman protocol. RF denotes radio frequency. S and D are the source and drain ohmic contacts of the single-electron transistor (SET) sensor, respectively. ST is the sensor’s top gate, and LB and RB are its barrier gates. C and R are the confinement and reservoir gates, respectively. P1 and P2 are plunger gates, and B1 and B2 are barrier gates. O marks the reservoir-side ohmic contact. Blue circles mark the sensor and qubit dots. The red arrow in the axis inset shows the in-plane field at angle $\phi$ from $\hat{x}$. (b) Example ESR spectrum far from the spin--valley anticrossing. Orange crosses and the solid orange curve show the data and fit, respectively. The blue and cyan dashed lines mark the intra-valley spin-flip transitions $A_1$ and $A_1'$. Imperfect initialization at the finite electron temperature of about $160~\mathrm{mK}$ can populate the first excited state, making $A_1'$ visible but weaker than $A_1$. (c) ESR spectra as a function of magnetic field, showing a pronounced anticrossing at $B=B_V$. The color scale shows spin-up signal in arbitrary units. The drive frequency $f_{\mathrm{MW}}=f_{\mathrm{LO}}+f_{\mathrm{IF}}$ combines the local-oscillator (LO) and intermediate (IF) frequencies. Orange dashed lines mark $B_V$ and the field in (b). The anticrossing occurs when the Zeeman energy matches the valley splitting $\Delta\approx66~\mu\mathrm{eV}$. The white double-headed arrow marks the splitting $2\lambda$, where $\lambda$ is the coupling strength extracted from the measurements.     
(d) Energy-level diagram of the four lowest spin–valley states $\{\ket{V1,\downarrow},\ket{V1,\uparrow},\ket{V2,\downarrow},\ket{V2,\uparrow}\}$, with colored arrows marking the observed transitions. For $B>B_V$, $A_1$ (blue) and $A_1'$ (cyan) are intra-valley ESR transitions in valleys V1 and V2, respectively, and $B_1$ (green) is an intervalley spin-conserving transition. The green replicas $B_1^{\mathrm{LO}}$ and $B_1^{\mathrm{LO-IF}}$ in (c) arise from residual mixer LO and sideband leakage, respectively. $C_2^\Lambda$ (magenta) corresponds to a two-photon process driven jointly by the main and leakage tones. $D_2^{\mathrm{1c}}$ and $D_2^{\mathrm{2c}}$ (red) are two-photon transitions involving either two main-drive photons or a mixed photon process, respectively.  
}

\label{fig:1}
\end{figure}

Figure~\ref{fig:1}(b) presents an ESR spectrum at B= 0.59 T. A doublet with a splitting of a few MHz can be seen. This indicates additional energy levels beyond a simple Zeeman-split two-level system.  To further investigate this observation, the ESR spectra are mapped as a function of magnetic-field applied along the DQD axis, and shown in Fig.~\ref{fig:1}(c). At each field, we held the LO frequency at $f_{LO}(B)=(27.7\,\mathrm{GHz/T})B-54.5\,\mathrm{MHz}$ and swept $f_{IF}$ from $10$ to $350\,\mathrm{MHz}$ in $1\,\mathrm{MHz}$ steps. At $B=0.578\,\mathrm{T}$ near the anticrossing, the main microwave tone $f_{MW}=f_{LO}+f_{IF}$ was swept from $15.9661$ to $16.3061\,\mathrm{GHz}$. The main feature is an anti-crossing that mixes a magnetic resonance inducing spin-flips that depends linearly on magnetic field, and an electric-dipole resonance inducing changes in an orbital degree of freedom that is independent of B-field. Hence, these levels arise from the presence of another orbital-like state with energy comparable to the Zeeman splitting. 

To explain the anti-crossing, we consider a single electron with four energy-levels labeled: $\ket{V1,\downarrow}$, $\ket{V1,\uparrow}$, $\ket{V2,\downarrow}$ and $\ket{V2,\uparrow}$. These are constructed from a spin up/down and two valleys, labeled $V1$ and $V2$, with energies $E_1$ and $E_2$ at zero B-field. We then construct an energy-level diagram versus B-field, which is depicted in Fig.~\ref{fig:1}(d). The curve $\left(\mathbf{A_1}\right)$ corresponds to the transition between the ground and the second excited state driven by the main microwave tone of the IQ-mixer at $f_{MW}=f_{LO}+f_{IF}$. In Fig.~\ref{fig:1}(c), when $B>B_V$, this gives a magnetic resonance linear in magnetic field, corresponding to the spin-flip transition within $V1$ manifold, i.e. $\ket{V1,\downarrow}\leftrightarrow\ket{V1,\uparrow}$, and when $B<B_V$ gives rise to an electric-dipole transition, independent of magnetic field, that flips the orbital  i.e. $\ket{V1,\downarrow}\leftrightarrow\ket{V2,\downarrow}$. The light blue line $\left(\mathbf{A_1'}\right)$ corresponds to analogous transition involving first and third excited states. The energy diagram has a mirror symmetry about the anti-crossing point, and consequently the transitions $\left(\mathbf{A_1,A_1'}\right)$ are nearly degenerate, except for a small difference in g-factor between $V1$ and $V2$. We assign the stronger ESR peak in  Fig.~\ref{fig:1}(b)  to the $V1$ transition, since we expect the Elzermann readout to preferentially load into the ground-state.    

Since confinement of the quantum dot is strongest along the $z$ direction (normal to the planar SOI substrate), the two lowest-energy states should correspond to the $\pm z$ valleys. Valley coupling at the $\mathrm{Si/SiO_2}$ interface lifts their twofold degeneracy,\cite{EDSRLeti} giving two spin-degenerate valley eigenstates $V1$ and $V2$ with energies $E_1$ and $E_2$, respectively. If the energy-splitting of $hf_V=66\,\mathrm{\mu eV}$ were due to an orbital splitting, the minimum possible length of the confinement potential, would be given by an infinite square potential of length $a_z\approx \sqrt{\frac{3\hbar^2\pi^2}{2m^*_t(E_2-E_1)}}=300\,\mathrm{nm}$ along the axis of the double quantum dot, where $m_t^*=0.2m_0$ is the effective mass of electron in transverse direction. This sets a minimum possible length for the confinement potential along the double quantum dot axis, and this is larger than the gate-pitch. Hence, we attribute the splitting to a valley, rather than an orbital splitting. Although the valley-splitting measured here is small for a planar Si-MOS device, it is similar to the valley-splitting measured in refs.\cite{GuoPhysRevLett.124.257701,tomić2025longcoherencesiliconspin}. The valley-orbit splitting can be reduced by steps in the $\mathrm{Si/SiO_2}$ interface\cite{PhysRevB.100.125309}. This suggests that the interface is not atomically flat over the extent of the electron wavefunction \cite{Cifuentes2024}.

Several additional resonances are also visible in the ESR spectra. They are labeled with colored letters corresponding to the transitions shown in Fig.~\ref{fig:1}(d), providing further confirmation of the energy-level structure. 
The vertical feint green line $\mathbf{B_1^\mathrm{LO}}$ is a replica of transition $\mathbf{B_1}$ driven by the  LO-leakage tone of the IQ-mixer. When making the map, the IF-frequency is fast scanned, and the LO-frequency is stepped with the magnetic field to track the magnetic transition; a vertical line arises at a B-field where $f_{LO}$ matches the $B_1$ transition. The diagonal green line $\mathbf{B_1^\mathrm{LO-IF}}$  with a high gradient occurs when the image sideband $f_{LO}-f_{IF}$ of the IQ-mixer matches the $\mathbf{B_1}$ transition. Although these leakage tones are $\sim40$ dB weaker than the main drive, the long $20\,\mathrm{\mu s}$ pulse duration allows sufficient time to coherently drive the transition $\mathbf{B_1}$. For $B<B_V$, the signal of magnetic resonance $\mathbf{B_1}$ degrades. When P1 gate is swept from operating to readout point, the system adiabatically passes through spin-valley anticrossing, swapping the $\ket{V1,\uparrow}$ and $\ket{V2,\downarrow}$ states, degrading the spin contrast. This is discussed in Supplementary Sec. “Elzerman readout with valley states”.

The purple line $\left(\mathbf{C_2^\mathrm{\Lambda}}\right)$ corresponds to a two-photon $\mathrm{\Lambda}$-transition observed when $B>B_V$. Here, a photon at $f_{LO}+f_{IF}$ first excites $\ket{V2,\downarrow}$ to a virtual state near $\ket{V2,\uparrow}$, and a photon at $f_{LO}$ then de-excites to $\ket{V1,\uparrow}$. The resonance occurs at $f_{IF}=f_Z-f_V$, where $f_Z$, $f_V$ are the Zeeman, and valley-splitting frequencies, respectively. Effectively, this two-photon $\Lambda$-transition acts as a net drive at $f_{IF}$. The visibility of $\mathbf{C_2^\mathrm{\Lambda}}$ increases when $f_{LO}\approx f_V$, since the virtual intermediate state approaches the real $\ket{V2,\uparrow}$ level.  We note that the signal strength of two-photon resonances is similar to the main one-photon resonances. A single quantum dot can be understood as a sensor of resonant electromagnetic fields that is easily saturated.  

The red lines $\left(\mathbf{D_2^\mathrm{1c}},\mathbf{D_2^\mathrm{2c}}\right)$ correspond with a two-photon process between states: $\ket{V1,\downarrow}\leftrightarrow\ket{V2,\uparrow}$. $\mathbf{D_2^\mathrm{1c}}$ is a one-color two-photon process driven by the main tone at $f_{LO}+f_{IF}$, and occurs midway between the two branches of the anti-crossing.  Counterintuitively, the two-color two-photon $\mathbf{D_2^\mathrm{2c}}$ resonance driven by photons at $f_{LO}+f_{IF}$ and $f_{LO}$ is stronger, despite the use of weak $f_{LO}$ tone. This is because the intermediate virtual state lies close to real states ($\ket{V1,\uparrow}$ or $\ket{V2,\downarrow}$), while the intermediate state of $\mathbf{D_2^\mathrm{1c}}$ is far detuned from any real states.  One-color two-photon resonances have previously been reported in EDSR experiments in Si/SiGe devices equipped with micromagnet\cite{PhysRevB.95.165429,PhysRevLett.115.106802}. A bichromatic two-photon resonance, as observed here, has been reported in $\mathrm{Ge/SiGe}$ hole spin devices\cite{PhysRevB.106.155412,PhysRevLett.132.067001}. There the bichromatic resonance was proposed as a method to scale up frequency multiplexing in a 2D array of quantum dots. Here however, the observed two-photon resonances illustrate potential cross-talk errors that may arise from the use of IQ-mixers as opposed to direct digital synthesis.

Together, the ESR measurements resolve all possible transitions accessible from the three lowest-energy states, providing direct experimental verification of the four-level spin–valley energy spectrum. Additional discussion of the spectra and readout is provided in the Supplementary Information.

\begin{figure}[h!]
\centering
\includegraphics[width=\linewidth]{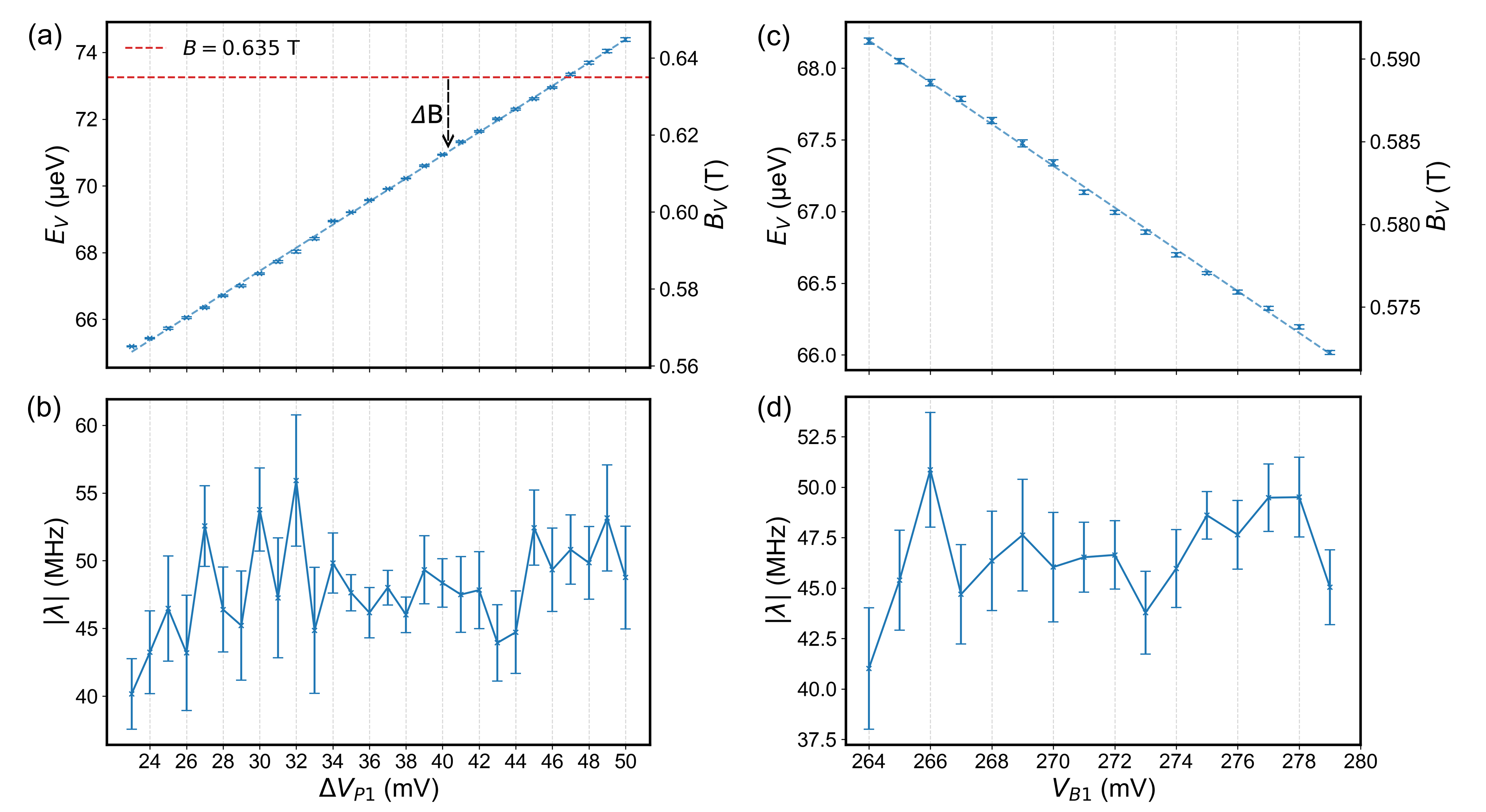}
\caption{Gate dependence of the valley splitting and inter-valley spin coupling. (a) Position of the anticrossing field $B_V$ and the corresponding valley splitting $E_V$ as a function of the plunger-gate voltage offset $\Delta V_{P1}$ from the resonant tunneling voltage. The blue dashed line is a linear fit. The red dashed line marks the applied magnetic field $B=0.635\,\mathrm{T}$, and the black arrow indicates the detuning $\Delta B=B-B_V$.
(b) Dependence of the inter-valley spin coupling magnitude $\lvert\lambda\rvert$ on $\Delta V_{P1}$.
(c) Position of the anticrossing field $B_V$ and the corresponding valley splitting $E_V$ as a function of the barrier-gate voltage $V_{B1}$. The blue dashed line is a linear fit. The linear dependence in (a) and (c) reflects the tunability of the valley splitting with vertical electric field and confinement electric potential.
(d) Dependence of $\lvert\lambda\rvert$ on $V_{B1}$. Error bars on the blue crosses show standard errors from the anticrossing fits.}
\label{fig:3}
\end{figure}

\subsection*{Gate-tunable valley splitting}

The valley-splitting is an interface interaction that is proportional to the overlap of the electron wavefunction with the interface, and hence the vertical electric-field \cite{PhysRevB.98.245424}. To test this, we measure the valley splitting $E_V\equiv E_{V2}-E_{V1}= g\mu_BB_V$ versus $\Delta V_{P1}$ with $V_{B1}=274\,\mathrm{mV}$, as shown in Fig.~\ref{fig:3}(a), and confirm a linear dependence, with $\frac{dE_V}{dV_{P1}}= 346.7\pm 1.5 \mathrm{\mu eV/V}$. $\Delta V_{P1}$ is the offset of the plunger-gate voltage from the resonant tunneling point $V_{P1}^{\mathrm{RES}}$, defined as the $P1$ gate voltage at the $(0\!\to\!1)$ charge transition. $\Delta V_{P1}=V_{P1}-V_{P1}^{\mathrm{RES}}$. This is consistent with previous reports in refs. ($\sim 600\,\mathrm{\mu eV/V}$)\cite{Yang2013,10.1063/1.5040474,10.1063/1.4972514}. The valley-splitting also decreases linearly with the barrier gate voltage with $\frac{dE_V}{dV_{B1}}=-145.6\pm 1.4 ~\mathrm{\mu eV/V}$ when $\Delta V_{P1}=30\,\mathrm{mV}$, see Fig.~\ref{fig:3}(c). This is compatible with a wavefunction that grows with barrier gate voltage.\cite{Goswami2007} The voltage control of the valley-splitting, supports the hypothesis that this is a valley rather than orbital state.\cite{PhysRevApplied.13.034068}

By comparison, the inter-valley spin coupling $\lambda$ is only weakly dependent on the gate-voltages, see Fig.~\ref{fig:3}(b,d) and hence $B_V$. This implies that $\lambda$ is relatively independent of magnetic field strength, and is dominated by the in-plane k-vector, rather than A-vector terms of the inter-valley spin-orbit interaction. 

\subsection*{Rabi frequency enhancement near anti-crossing}

In Fig.~\ref{fig:1}(c) the anti-crossing involves a diagonal line corresponding to a magnetic resonance that flips spin, and a horizontal line corresponding to a flip in the valley-orbital degree of freedom driven by an electrical field. Consequently, near the anti-crossing the Rabi frequency should be sensitive to the admixing caused by the spin-valley coupling\cite{PhysRevB.97.155433}. We note that in EDSR experiments on Si-FDSOI corner dot devices with micromagnet\cite{Klemt2023}, and without micromagnet\cite{EDSRLeti} an enhancement in Rabi frequency near the valley anti-crossing has been reported. Furthermore, EDSR resonances have been reported in similar Si-MOS devices \cite{tomić2025longcoherencesiliconspin}. Here we explore if an enhancement in Rabi frequency can be observed in planar Si-MOS device. 
 
\begin{figure}[h!]
\centering
\includegraphics[width=\linewidth]{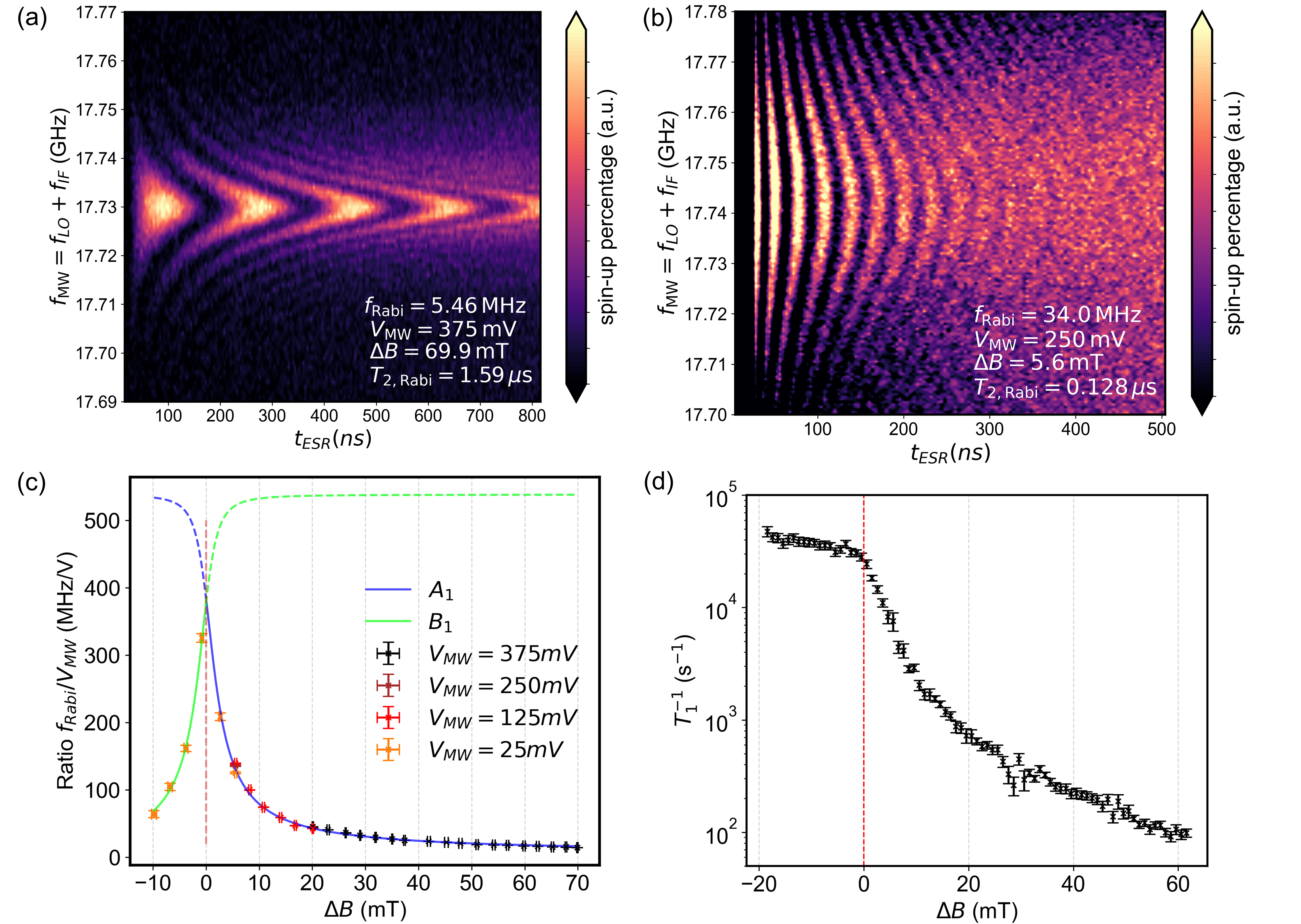}
\caption{Rabi oscillations and energy relaxation near the spin--valley anticrossing. (a) Rabi chevron at detuning $\Delta B=69.92\pm0.15\,\mathrm{mT}$ and microwave source voltage $V_{\mathrm{MW}}=375\,\mathrm{mV}$, with Rabi frequency $f_{\mathrm{Rabi}}=5.46\pm0.01\,\mathrm{MHz}$.
(b) Rabi chevron at $\Delta B=5.6\pm0.2\,\mathrm{mT}$ and $V_{\mathrm{MW}}=250\,\mathrm{mV}$, with $f_{\mathrm{Rabi}}=34.0\pm0.3\,\mathrm{MHz}$. In (a) and (b), the color scales show the spin-up probability. Here $t_{\mathrm{ESR}}$ is the microwave pulse duration, $f_{\mathrm{MW}}=f_{\mathrm{LO}}+f_{\mathrm{IF}}$ is the sum of the local oscillator and intermediate frequencies, and $T_{2,\mathrm{Rabi}}$ is the Rabi damping time.
(c) Ratio of Rabi frequency to applied microwave source voltage as a function of detuning $\Delta B=0.635\,\mathrm{T}-B_V$, where $B_V$ is the anticrossing field. Black, brown, red, and orange crosses correspond to $V_{\mathrm{MW}}=375$, $250$, $125$, and $25\,\mathrm{mV}$, respectively. The blue and green curves are fits to the $A_1$ and $B_1$ branches, with dashed extensions across the anticrossing. Horizontal error bars give the detuning uncertainty from the valley-splitting calibration, and vertical error bars give the normalized Rabi fit standard errors. The red dashed line marks zero detuning. An enhancement is observed near the anticrossing, exceeding a factor of 20 relative to the far-detuned case. At $\Delta B=2.7\,\mathrm{mT}$, a lower drive is needed to time-resolve the high-speed Rabi oscillation.
(d) Measured energy relaxation rates $1/T_1$ at $\Delta V_{P1}=30\,\mathrm{mV}$, where $\Delta V_{P1}=0$ is the resonant tunneling condition, plotted against detuning $\Delta B=B-B_V$, where $B_V=578\,\mathrm{mT}$. The black crosses are extracted from exponential fits to the decay of the blip probability with wait time, as described in the text. Error bars show the corresponding fitted standard errors. The red dashed line marks zero detuning.}

\label{fig:rabi}
\end{figure}
To characterize this effect, we measure the Rabi frequency as a function of detuning from the anticrossing. The measurement is made at fixed external magnetic field of $B=0.635\ \text{T}$ along the $\hat{x}$, $[110]$ direction. This allows the use of a nearly constant microwave frequency to maintain a consistent microwave power independent of frequency dependent power delivery due to standing waves etc. 
The detuning $\Delta B = 0.635\ \text{T} - B_V$ is controlled via $V_{P1}$, which linearly shifts $B_V$ (Fig.~\ref{fig:3}(a)), so that positive (negative) values correspond to operation above (below) the anticrossing. Fig.~\ref{fig:rabi}(a,b) show examples of Rabi chevrons driven via the ESR antenna at two different detunings far and near the anti-crossing, respectively. Near the anti-crossing, the Rabi frequency seen in Fig.~\ref{fig:rabi}(b) is x6 faster than the far from anti-crossing case, see Fig.~\ref{fig:rabi}(a), despite the lower microwave drive.

Figure \ref{fig:rabi}(c) plots the ratio of the Rabi frequency to the microwave source voltage $V_{MW}$ versus the detuning. The data are collected for a few different microwave drives, which are plotted in different colors. Near the anti-crossing, the measurement is limited to a bandwidth of 125 MHz by the time-resolution of the arbitrary waveform generator, and decreasing the Rabi frequency helps to measure in this regime. Far from the anti-crossing, using a larger drive allows a faster Rabi frequency, and higher Q-factor, which allows easier measurement of Rabi frequency. This analysis assumes the Rabi frequency is proportional to the voltage amplitude of the IF-signal applied to the IQ-mixer, $V_{MW}$.    

What is striking is that the ratio of Rabi frequency to microwave source voltage increases by a factor of more than 20 near the anti-crossing. This is a bit unexpected, since the antenna is optimized for magnetic field generation. For a transition driven by an a.c. magnetic field, the magnetic-drive contribution to the Rabi frequency between the ground and first excited state (transition B) is $hf_{B,12}= \,\frac{1}{2}g\mu_B B^{\mathrm{eff}}_{\perp}\bra{2}(\ket{\uparrow}\bra{\downarrow}+\ket{\downarrow}\bra{\uparrow})\ket{1}\,$, where $\ket{1} \approx \ket{V1,\downarrow}$, and $\ket{2} \approx a_2\ket{V1,\uparrow} +b_2\ket{V2,\downarrow}$, where $B^{\mathrm{eff}}_{\perp}$ is the effective amplitude of the oscillating magnetic field perpendicular to the static field along the device channel $[110]$. This model assumes a rotating-wave-approximation where only the two levels on resonance with the drive are important. For a spin-flip magnetic resonance only, the Rabi frequency would decrease near the anti-crossing. However, an enhancement is observed. 

To explain the enhancement, we consider an additional electric-dipole term that flips the valley degree of freedom: $hf_{E,12}=\,\mu_D E_{\mathrm{ac}}\bra{2}(\ket{V1}\bra{V2} + \ket{V2}\bra{V1})\ket{1}\,$, where $E_{\mathrm{ac}}$ is the amplitude of the oscillating electric field at the dot and $\mu_D$ is the inter-valley dipole matrix element. Capacitive coupling from the antenna to nearby gates is a possible source of this drive \cite{Dehollain2013}. This results in a total Rabi frequency $f_{12}=\vert f_{E,12}+f_{B,12}\vert $. From fit to data of Fig.~\ref{fig:rabi}(c) we extract an effective inter-valley coupling $|\lambda_{\mathrm{eff}}|=39.39\pm0.97\,\mathrm{MHz}$, $|\mu_D E_{\mathrm{ac}}|=538\pm8\,\mathrm{MHz}$ and $|1/2g\mu_B B^{\mathrm{eff}}_{\perp}|=5.56\pm0.76\,\mathrm{MHz}$ at an applied microwave drive of $V_{MW}=1\,\mathrm{V}$. The ratio of the fitted electric and magnetic drive prefactors is $\frac{|\mu_D E_{\mathrm{ac}}|}{|\tfrac{1}{2}g\mu_B B^{\mathrm{eff}}_{\perp}|}\approx 100$.

Using \(17\,\mathrm{GHz}\) as a representative frequency within the experimental range, an electromagnetic simulation of the antenna and gate layout gives \(|E_z|/|B_y|\simeq34.6\,\mathrm{kV\,m^{-1}\,mT^{-1}}\) at the modelled P1 dot position, see Supplementary Fig.~\ref{fig:hfss_17ghz}, where \(E_z\) is the vertical electric field and \(B_y\) and \(B_z\) are the transverse magnetic field components. Assuming a vertical inter-valley dipole in the flat-interface approximation, we take $E_{\mathrm{ac}}=|E_z|$ and obtain\(|\bra{V_2}r_D\ket{V_1}|=\frac{|\mu_D E_{\mathrm{ac}}|}{e|E_z|}\approx 0.16\,\mathrm{nm}.\)

Spin-valley mixing near the anti-crossing enables faster electrically driven control. However, the Rabi damping times are shorter, due to a Rabi frequency that is more sensitive to charge noise, consequently the best Q-factors are observed away from the anti-crossing, see Supplementary Fig.~\ref{fig:Q-factor}.

\subsection*{Relaxation hot-spot measurement}

The Rabi frequency enhancement arises from admixing due to inter-valley spin coupling. This can also give rise to a spin relaxation hot-spot \cite{Yang2013,Huang_PhysRevB.90.235315}. To confirm this, we measure the relaxation time $T_1$ in the vicinity of the anti-crossing using a three-stage pulsing protocol applied to the plunger gate P1. The sequence consists of:
(i) an \emph{empty stage}, where the quantum dot is pulsed into the $N=0$ charge state;
(ii) a \emph{load stage}, where the dot is pulsed into the $N=1$ regime such that an electron is randomly loaded into one of the four spin–valley states; and
(iii) a \emph{readout stage}, where the dot is pulsed to a read-out point close to the resonant tunneling condition between the $N=0$ and $N=1$ charge states, enabling energy-selective tunneling. In the case of a two-level system, this would result in a spin-selective readout\cite{Elzerman2004,KeithElzermann}. Here however, there are four energy-levels and the thermal energy $k_BT_e\simeq14~\mu\mathrm{eV}$ at $163~\mathrm{mK}$ is about a quarter of the valley splitting $E_V\approx60~\mu\mathrm{eV}$ and the Zeeman energy near the anti-crossing, so the interpretation of the readout is more involved. Supplementary Fig.~\ref{fig:readout_p1} shows their alignment with the reservoir at the read-out point.

To extract $T_1$, we vary the waiting time during the load stage. Since the electron can initially occupy any of the four spin–valley states, relaxation processes during this waiting period repopulate the ground state. To probe the probability of remaining in an excited state, we measure the blip fraction as a function of wait time and fit it to a single exponential decay to extract $T_1$. This gives an effective relaxation time because the states can be populated and detected with different probabilities. For all measurements, the read-out point is a fixed voltage-shift from the resonant tunneling condition, and the same cut-off time is used.

The relaxation time measurement is performed at $\Delta V_{P1}=30$ mV for different magnetic fields across the fixed anticrossing. At this gate voltage, the anticrossing occurs at $B_V \approx 578$ mT. We therefore define the detuning variable as $\Delta B = B - 0.578~\text{T}$, which serves as the independent parameter in Fig.~\ref{fig:rabi}(d).

In Yang {\it{et al.}}'s work\cite{Yang2013}, a hot-spot in the spin-relaxation is observed on top of a $B^5$ dependence. Here however, for $B<B_V$, the relaxation rate plateaus at a fast value. In Yang {\it{et al.}}'s work\cite{Yang2013}, the interpretation assumes that inter-valley relaxation is too fast to measure, and what is observed is slower spin relaxation of the mostly $\ket{V1,\uparrow}$ state %(us 10ms to $<$ 100us, are compatible with their measurements) 
due to phonon-mediated spin-flip, and a spin-valley flip-flop process that gives rise to a resonant relaxation process.

To interpret our experiment, we first note that the Elzerman readout is an energy-selective process and distinguishes between high (the three excited states) and low (the ground state) energy states at the readout point. We label the states 1-4, from low to high energy. For $B > B_V$, the measurement mostly discriminates the spin states due to fast relaxation of valley states. The excited valley state 4 quickly relaxes predominantly to state 3, $\ket{V2,\uparrow}\to\sim\!\ket{V1,\uparrow}$, with some relaxation to state 2, and state 2 quickly relaxes to state 1, $\sim\!\ket{V2,\downarrow}\to\ket{V1,\downarrow}$. Hence, the observed energy relaxation $T_1$ time is dominated by the slower relaxation of state 3. Far from the anti-crossing, the relaxation is dominated by spin-flip, but near the anti-crossing the relaxation rate increases due to a valley-flip activated by the admixing of state 3, as seen in Fig.~\ref{fig:rabi}(d).

For $B < B_V$, we note that the valley splitting depends on the P1 gate voltage, see Fig.~\ref{fig:3}(a), and when the P1 gate voltage is swept from the wait to read-out point, it adiabatically sweeps through the anti-crossing of states 2 and 3, see Supplementary Fig.~\ref{fig:readout_p1}. On read-out, state 3 maps from an excited valley to long lived ground valley state as  $\sim\!\ket{V2,\downarrow}\to\sim\!\ket{V1,\uparrow}$, resulting in a blip. State 2 maps to an excited valley-state, $\sim\!\ket{V1,\uparrow}\to\sim\!\ket{V2,\downarrow}$ that can relax via a valley flip before a blip is detected, $\ket{V2,\downarrow}\to\ket{V1,\downarrow}$ . Consequently, for $B < B_V$ the valley relaxation of states  $3\to1$ is predominantly observed, resulting in the plateau in the energy relaxation rate seen in Fig.~\ref{fig:rabi}(d). This supports our understanding of the Rabi frequency enhancement.

\subsection*{g-factor anisotropy for the V1 and V2 valleys}

\begin{figure}[h!]
\centering
\includegraphics[width=\linewidth]{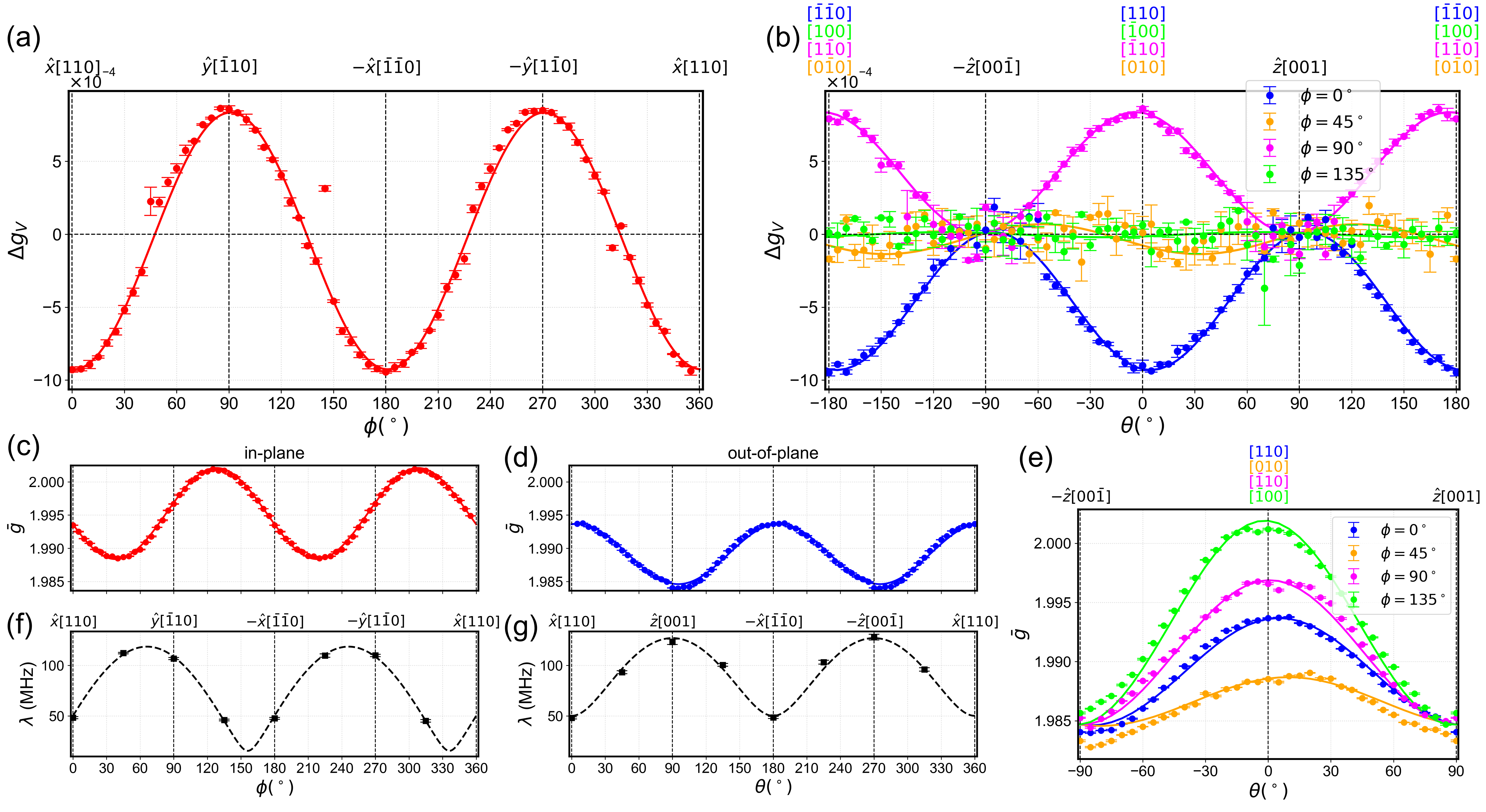}
\caption{Angular dependence of the valley $g$-factors and inter-valley spin coupling. (a) Anisotropy of the difference in $g$-factor between valley states, $\Delta g_V=\lvert g_{V1}\rvert-\lvert g_{V2}\rvert$, for an in-plane magnetic field. The angle $\phi$ is measured from $[110]$. The $\Delta g_V$ anisotropy is aligned along the $[110]$ and $[\bar{1}10]$ crystal axes.
(b) $\Delta g_V$ for out-of-plane sweeps at fixed $\phi$. The angle $\theta$ is measured from the in-plane direction, with $\theta=90^\circ$ along $[001]$.
(c) Anisotropy of the mean $g$-factor, $\bar{g}=(\lvert g_{V1}\rvert+\lvert g_{V2}\rvert)/2$, for an in-plane magnetic field. The in-plane minimum and maximum are nearly aligned with the $[010]$ and $[100]$ crystal axes, respectively.
(d) $\bar{g}$ for an out-of-plane sweep in the $xz$ plane.
(e) $\bar{g}$ for out-of-plane sweeps at four values of $\phi$. In (b) and (e), blue, orange, magenta, and green correspond to $\phi=0^\circ$, $45^\circ$, $90^\circ$, and $135^\circ$, respectively.
(f) Anisotropy of the inter-valley spin coupling $\lambda$ measured from the anticrossing for an in-plane magnetic field.
(g) $\lambda$ for an out-of-plane sweep in the $xz$ plane. The anisotropy of $\lambda$ is consistent with a spin--valley field that lies in the plane.
Circles in (a--e) and black squares in (f) and (g) show the data. Solid curves in (a--e) and black dashed curves in (f) and (g) show the fits. Vertical dashed lines mark the indicated crystal directions, and horizontal dashed lines in (a) and (b) mark $\Delta g_V=0$. Error bars show the uncertainties in the plotted quantities. In (c--e), each error bar shows the average of the uncertainties in $g_{V1}$ and $g_{V2}$.}

\label{fig:2}
\end{figure}

To understand the symmetry of the spin-valley coupling, we start by probing the intra-valley spin coupling via the g-factor anisotropy using a magnetic field far from the anti-crossing. In Fig.~\ref{fig:2}(a), we define the magnetic field orientation using spherical coordinates. As shown in Fig.~\ref{fig:1}(a), the in-plane angle $\phi=0$ corresponds to the positive $\hat{x}$ direction, which aligns with both the main axis of the device and the vector-magnet coils, along the $[110]$ crystal direction. The out-of-plane angle is defined such that $\theta=0$ lies in the device plane, while $\theta=90^\circ$ corresponds to the field pointing along the positive $\hat{z}$ axis, i.e. the [$001$] direction.

To extract the valley-dependent $g$-factors, we perform angular-dependent ESR measurements. For each magnetic-field orientation $(\theta,\phi)$, the $g$-factor is obtained from measurements of the ESR frequencies for opposite field directions, $B$ and $-B$, at B-field magnitudes around $B=635\,$mT, far from the anti-crossing. This approach exploits the intrinsic symmetry between opposite field polarities and provides a robust method to determine the angular dependence of the $g$-factor in both valleys.

We start by discussing the anisotropy of the difference in magnitude of g-factor between V1 and V2 states, $\Delta g_V= \vert g_{V1}(\theta,\phi)\vert-\vert g_{V2}(\theta,\phi)\vert$. This is measured using the splitting between V1 and V2 magnetic resonances, an example data set is shown in Fig.~\ref{fig:1}(b). As far as we are aware, the difference in g-factor between valley-states for single quantum dot has only been reported previously in refs. \cite{Ferdous2018,tomić2025longcoherencesiliconspin}. Ferdous {\it{et al.}}\cite{Ferdous2018} measure for Si/SiGe quantum dots equipped with a micromagnet. Here, there is no micromagnet, resulting in a more direct measurement without the need to account for the anisotropy of the B-field of the micromagnet. Tomic {\it et al}\cite{tomić2025longcoherencesiliconspin} measure the difference in g-factor between two nearby quantum dots for similar Si-MOS devices as reported here. Measurements for the case where both electrons are in the ground-valley, and the case where one of the electrons is in the excited valley are reported. From this data, it is possible to deduce the difference in g-factor between the valleys of one of the quantum dots, but the answer is multi-valued due to sign ambiguities. Here a more direct measurement of the difference in g-factor magnitude between V1 and V2 states of single quantum dot is reported.

In Fig.~\ref{fig:2}(a), $\Delta g_V$ is measured as a function of in-plane magnetic field. It varies sinusoidally as $\Delta g_V\frac{\mu_BB}{h} = B_{xy} (\Delta \alpha_{Vg}+\Delta \beta_{Vg} \sin{(2\phi')})$, where $B_{xy}$ is the in-plane component of the B-field and $\phi'=\phi+\pi/4$ is the in-plane angle measured from $[100]$. The Rashba term $|\Delta \alpha_{Vg}| = 0.7\pm0.3\,$MHz/T, and Dresselhaus term $|\Delta \beta_{Vg}| =19.8\pm0.4\,$MHz/T. The minimum/maximum are aligned with the DQD axes $[110]$, and $[\bar{1}10]$, respectively. In Fig.~\ref{fig:2}(b), the out-of-plane dependence of $\Delta g_V$ is measured for various in-plane angles $\phi$. The data are consistent with a model in which only the in-plane B-field component is important. The anisotropy in the difference in g-factor arises from the vector-potential component of the intra-valley spin coupling. The alignment of $\Delta g_V$ with $[110]/[\bar{1}10]$ crystal-axes implies that $g_{[100],[010]}^{V1}=-g_{[100],[010]}^{V2}$. This is consistent with  the theory of Ruskov {\it{et al.}}\cite{Ferdous2018,PhysRevB.98.245424,PhysRevB.92.201401,PhysRevB.97.241401}, see Supplementary Sec. \ref{sec:Delta_g}, where
the intra-valley Dresselhaus coefficients $\beta_{V1}=-\beta_{V2}$, and are much larger than the Rashba terms.

Next, we discuss the anisotropy of the mean magnitude of the $g$-factor of the two valley states, $\bar{g}_V=(\vert g_{V1}(\theta,\phi)\vert+\vert g_{V2}(\theta,\phi)\vert)/2$. We note that the modulation in mean g-factor is on the order of $1\%$. This is too large to be due to systematic calibration error in the vector magnet, and the measured g-factor is not aligned with the magnet axes. Hence, we attribute the variation to the device. The maximum in mean g-factor is aligned in-plane at nearly $45^{\circ}$ with respect to the DQD axis $[110]$, along the $[100]$ crystal-axis. This is also consistent with a model where in the $([100],[010])$ basis $g_{V1}$ and $g_{V2}$ differ in the sign of the off-diagonal terms only, $g_{[100],[010]}^{V1}=-g_{[100],[010]}^{V2}$, see Supplementary Sec. \ref{sec:mean_g}. 

The mean magnitude of the g-factor is found from a global fit to the data to be
\begin{equation}
\mathbf{\bar{g}}_{V}
=
1.98468(5)\,\mathbb{I}
+
\begin{bmatrix}
0.891(8) & -0.666(5) & 0.081(6) \\
-0.666(5) & 1.219(8) & 0.015(6) \\
0.081(6) & 0.015(6) & 0
\end{bmatrix}
\times 10^{-2}.
\end{equation}
Here, $\mathbb{I}$ is the $3\times3$ identity matrix, and the axes follow the order $\{\hat{x},\hat{y},\hat{z}\}$.

This is different to observations of other studies\cite{gfactorexp}, where the mean g-factor of single quantum dot is reported as nearly aligned with $[110]$ axis. We further note, that there are a number of reports of the difference in g-factor between two nearby quantum dots where $\Delta g_{DQD}^{V1}$ is also aligned with $[110]/[\bar{1}10]$ crystal axes\cite{Jock2018,gfactorexp,Cifuentes2024,chittockwood2025radiofrequencycascadereadoutcoupled}. In principle, the anisotropy of the inter-valley spin coupling could give rise to an anisotropy in the mean g-factor due to a small shift in energy of  the admixed energy level of the ESR transition far from the anti-crossing. However, this effect is too small to explain the observed anisotropy. Also, we note that if the aluminium antenna is in a superconducting state, the Meissner effect could screen the external magnetic field resulting in an anisotropy in the g-factor, as noted in  ref. \cite{Kerckhoff_PRXQuantum.2.010347}. However, we rule this out as here the magnetic field of $\sim 600~\mathrm{mT}$ is relatively high compared to the critical B-field of bulk aluminium of $B_c\sim 10~\mathrm{mT}$ \cite{Caplan_PhysRev.138.A1428}, and for a $\approx 100~\mathrm{nm}$ thick stripline, the enhancement in the critical in-plane magnetic field should be insufficient to remain superconducting \cite{Reale}.

\subsection*{Anisotropy of inter-valley spin coupling}

We investigate the anisotropy of the spin–valley coupling strength by performing angular-dependent measurements of the anticrossing map. From a fit to the anti-crossing, a value $B_V$ corresponding to the B-field magnitude at which the crossing occurs, and an inter-valley spin coupling strength $\lambda$. The zero-field valley-splitting, proportional to $B_V$ is found to be independent of the B-field orientation. The results for spin-valley coupling $\lambda$ are shown in Figs.~\ref{fig:2}(f) and \ref{fig:2}(g) for magnetic-field rotations in the $\hat{x}$–$\hat{y}$ and $\hat{x}$–$\hat{z}$ planes, respectively.

The inter-valley spin coupling acts as an effective magnetic field $B_{VSO}$ that points in the plane of the Si/$\mathrm{SiO}_2$ interface, since $H_{VSO}=(\alpha_{VSO}-\beta_{VSO})k_{y}\,\sigma_{x}-(\alpha_{VSO}+\beta_{VSO})k_{x}\,\sigma_{y} \equiv \mu_B\vec{b}_{VSO}\cdot\boldsymbol{\sigma}$. The inter-valley spin coupling $\lambda$ is proportional to the component of $\vec{b}_{VSO}$ that is transverse to the external magnetic field direction $\hat{B}_{ext}$, $\lambda = \mu_B \vert \vec{b}_{VSO} \times \hat{B}_{ext}\vert $, which is a maximum for out-of-plane B-field, see Fig.~\ref{fig:2}(g). Furthermore, in Fig.~\ref{fig:2}(f) $\lambda $ does not go to zero for any in-plane magnetic field, implying that the valley spin-orbit field $\vec{b}_{VSO}$ is complex\cite{GuoPhysRevLett.124.257701}.
The maximum/minimum points $\lambda$ are misaligned with respect to the crystal and device axes. This can occur if $\langle k_x^2\rangle \neq \langle k_y^2\rangle$, and $\alpha_{VSO}\neq 0$. This is to be expected, as the wavefunction will be elongated along the axis of DQD channel.

We fit the dataset to the model, see Supplementary Sec. \ref{sec:lambda},
\begin{equation}
\lambda^2(\phi,\theta; b,A_{\lambda},\phi_{0},\theta_{0})
=
(\mu_Bb)^{2}\left[
\sin^{2}(\theta-\theta_{0})
+
\frac{1}{2}\cos^{2}(\theta-\theta_{0})
\left(1-A_{\lambda}\cos(2\phi-\phi_{0})\right)
\right] .
\end{equation}

with best global fit parameters: $\mu_Bb=124\pm 1 ~\mathrm{MHz}$, $A_{\lambda}=0.91\pm0.05$, $\phi_0 = -47\pm 3^{\circ}$, $\theta_0 = 0\pm 2^{\circ}$.

Previously, the inter spin-valley coupling anisotropy has been measured indirectly via rate of spin relaxation. Zhang {\it{et al.}}\cite{GuoPhysRevLett.124.257701} measure anisotropy of $T_1$ times in planar Si-MOS device, and find minimum rate when B-field is slightly offset from $[010]$ direction, at $45^\circ$ to the DQD axis. In work of Spence {\it{et al.}}\cite{PhysRevApplied.17.034047} on a Si-CMOS nanowire with corner dots, $T_1$ spin relaxation rates are measured as a function of out-of-plane B-field. The $T_1$ relaxation rates were found to be invariant for a rotation about the nanowire axis $[110]$, indicating a spin-orbit field $B_{VSO}$ aligned along $[110]$. Here, the nanowire channel is wider, and the $B_{VSO}$ is rotated away from $[110]$.   

\section*{Discussion}

We have presented an example of a planar Si-MOS quantum dot with a relatively small valley splitting of about $66~\mathrm{\mu eV}$. Usually small valley-splittings are associated with Si/SiGe quantum wells, where the potential step at the interface is less steep than the case of $\mathrm{Si/SiO_2}$ interfaces. Other works have also reported the occasional small valley splitting\cite{GuoPhysRevLett.124.257701}. Electron-spin resonance experiments near the valley anti-crossing are performed. A vector-magnet is used to study the anisotropy of the g-factor and inter-valley spin coupling strength. The difference in g-factor between the valleys is aligned with $[110]/[\bar{1}10]$ crystal-axes. The mean g-factor varies on the 1\% level, and is aligned along axis close to $[100]/[010]$ in the plane. This is consistent with a $g_{[100],[010]}^{V1}=-g_{[100],[010]}^{V2}$, which is expected when the Dresselhaus term dominates the intra-valley spin-coupling \cite{PhysRevB.98.245424,Ferdous2018}. From the anti-crossing, we can measure the inter-valley spin coupling strength, and find an effective spin-valley field that lies in the plane at an angle that is not aligned with the crystal-axes. This is consistent with an interface spin-orbit effect with a wavefunction where the envelope is elongated $\langle k_x^2\rangle \neq \langle k_y^2\rangle$. Near the anti-crossing the relaxation rate increases on approaching from higher field, consistent with enhanced spin relaxation due to inter-valley spin coupling. 

Of most interest is the observation of an enhancement in the Rabi frequency near the anti-crossing. This is attributed to the activation of an electric-dipole transition that flips the valley degree of freedom, with a fitted drive prefactor about 100 times that of the magnetic drive. We speculate that in addition to generating an alternating magnetic field, driving the antenna also generates an electric-field, possibly due to capacitive coupling with either the confinement or plunger gate. Usually, for electron spin in silicon, observation of electric-dipole spin resonance uses a micromagnet\cite{Klemt2023}, or a highly non-planar device\cite{Jock2018}, so it is noteworthy that EDSR in planar Si-MOS quantum dots may be viable route for fast qubit control. This may take the form of two-photon control schemes at B-fields far from the anti-crossing. 

Finally, the valley-states are an important consideration for shuttling\cite{Volmer2024,PRXQuantum.4.020305}, spin-readout\cite{Philips2022}, and spin dephasing due to electrical noise\cite{PhysRevB.107.085427}. This work presents characterization of spin-valley coupling in a planar Si-MOS device, and contributes to an understanding of this important device property.

\section*{Methods}

The devices are natural Si-MOS double quantum dots on a silicon-on-insulator wafer from the same batch as ref. \cite{peetroons2025highfidelityqubitcontrol}. The devices are equipped with a single electron transistor for charge detection and an antenna for ESR. The devices are fabricated at IMEC, and further details on similar devices can be found in refs. \cite{Elsayed2024,LiIMEC}.

The single electron transistor (SET) consists of a large quantum dot formed between gates LB and RB which are underneath gate ST. In the silicon channel defined by confinement gate C (channel width $26\,$nm), up to two smaller quantum dots can be formed in series beneath plunger gates P1 and P2 (gate width $63\,$nm, gate pitch $58\,$nm). The plunger gate layer is deposited above the confinement gate C, while barrier gates B1, B2, and B3 (gate width $22\,$nm, pitch $32\,$nm) sit between and above the plunger gates and are used to tune the tunnel barriers of the two smaller quantum dots.  

We employ radio-frequency (RF) reflectometry for fast and sensitive charge readout. The tank circuit, consists of a $0.7\,$pF coupling capacitor and a $270\,$nH inductor connected to the S contact of the SET in parallel \cite{RFreview}. Measurements are performed in a dry dilution refrigerator with a base temperature of $80\,$mK. With an integration time of $10\,\mu$s, the ratio of the change in signal between on and off the SET Coulomb peak and the noise is 80.

The SET is used to sense the charge state of the quantum dot beneath gate P1. This dot is depleted down to the last electron. The spin state is detected using the Elzerman spin-to-charge conversion method \cite{KeithElzermann}. A reservoir is formed adjacent to the P1 dot by biasing gates P2, B2, and R to $3\,$V, well above the gate threshold voltage of $\sim 0.3\,$V. Gate B1 is tuned to control the tunnel rate between the P1 dot and the reservoir.  

For the ESR experiments, we perform single-shot spin readout using a two-stage pulsing protocol on the plunger gate P1. The sequence consists of:  
(i) a \emph{load (operation) stage}, where the dot is pulsed into the $N=1$ charge state so that a single electron is loaded into the quantum dot and where the MW pulses are applied; and  
(ii) a \emph{readout (empty) stage}, where the dot is pulsed to the calibrated read-out point near the boundary between $N=0$ and $N=1$ see Supplementary Fig.~\ref{fig:readout_p1}.

During the readout stage, energy-dependent tunneling occurs. An electron in an excited state is more likely to tunnel to the reservoir, briefly emptying the dot before reloading. This process maps the spin–valley state onto a detectable change in the charge occupancy. A time trace of the SET Coulomb peak signal is recorded, allowing discrimination between $N=1$ and $N=0$ charge states. The signal-to-noise ratio (SNR) for this single-shot readout is about 20.

\bibliography{sample}

\section*{Data Availability}
The data that support the findings of this study are openly available in Zenodo at {\urlstyle{same}\url{https://doi.org/10.5281/zenodo.24344814}}\cite{LuoValleyData}.

\section*{Acknowledgements}
Not applicable.

\section*{Funding}
This work was financially supported by the JST Moonshot R\&D under Grant Nos. JPMJMS2065 and JPMJMS256H and the Engineering and Physical Sciences Research Council (EPSRC) under Grant No. EP/S022953/1. X.L. acknowledges support from the Cambridge Trust through a Cambridge International Scholarship. X.P. acknowledges the University of Cambridge Harding Distinguished Postgraduate Scholars Programme. The funders had no role in study design, data collection and analysis, preparation of the manuscript, or the decision to submit it for publication.

\section*{Competing Interests}
The authors declare no competing financial or non-financial interests.

\section*{Author contributions statement}

\section*{Additional information}

\clearpage
\section*{Supplementary Information for: Valley enhanced Rabi frequency in n-type planar Silicon-MOS quantum dot}

\medskip
\setcounter{figure}{0}
\renewcommand{\theHfigure}{supplement.\arabic{figure}}
\captionsetup[figure]{name=Supplementary Figure}

\renewcommand{\thesection}{\Roman{section}}
\section{Readout calibration}
\label{sec:readout_calibration}
At each B-field, we locate the resonant tunnelling voltage $V_{P1}^{\mathrm{RES}}$  by tuning P1 until the dot spent equal time in empty and singly occupied states. At this point, the lowest energy-level of the singly charged quantum dot is approximately level with the reservoir. We set P1 $300\,\mu\mathrm{V}$ above this voltage for readout on both sides of $B_V$. This was determined from an Elzerman spin-tail measurement, where an electron is randomly loaded into one of the 4 spin-valley states. At this voltage, state 2 and 3 is close to resonance with the reservoir. The P1 lever arm is $\alpha_{P1} =0.215\,\mathrm{eV/V}$. We allowed $10\,\mu\mathrm{s}$ for settling due to the $100\,\mathrm{kHz}$ low-pass filter on the gates. The integration time is $10\,\mu\mathrm{s}$, and the barrier is set to achieve an average blip-time of $100\,\mu\mathrm{s}$, and the recording window is set to $700\,\mu\mathrm{s}$, following recommendations of Keith\citesupp{si-Keith_2019}.

\section{Elzerman readout with valley states}

To better understand Elzerman readout in presence of valley states, we plot the energies of the four spin-valley levels with respect to the ground state as a function of P1 gate voltage in Supplementary Fig.~\ref{fig:readout_p1} , using data of Fig.~\ref{fig:3}. The states are labelled 1--4, from low to high energy. On read-out, the P1 gate voltage is adiabatically swept from the operating point at $\Delta V_{P1}=+30\,\mathrm{mV}$, to the readout point at $\Delta V_{P1}=+0.3\,\mathrm{mV}$. For $B>B_V$, the states are unchanged, and blips occur for higher energy-states. For the excited valley-states, the fast valley relaxation will limit the time to detect a blip, and the blip probability is highest for $\ket{V1,\uparrow}$, resulting in a measurement with high spin contrast. For $B<B_V$, states 2 and 3 anticross, such that the state with high blip probability was $\ket{V2,\downarrow}$ at the operating point. This results in a measurement with a high valley contrast. Hence in Fig.~\ref{fig:rabi}(d), the valley relaxation rate is measured for $B<B_V$, and spin-relaxation for $B>B_V$. In Fig.~\ref{fig:1}(c), the resonances result in a change in the energy of the electron and a contrast in signal.

\begin{figure}[h!]
\centering
\includegraphics[width=\linewidth]{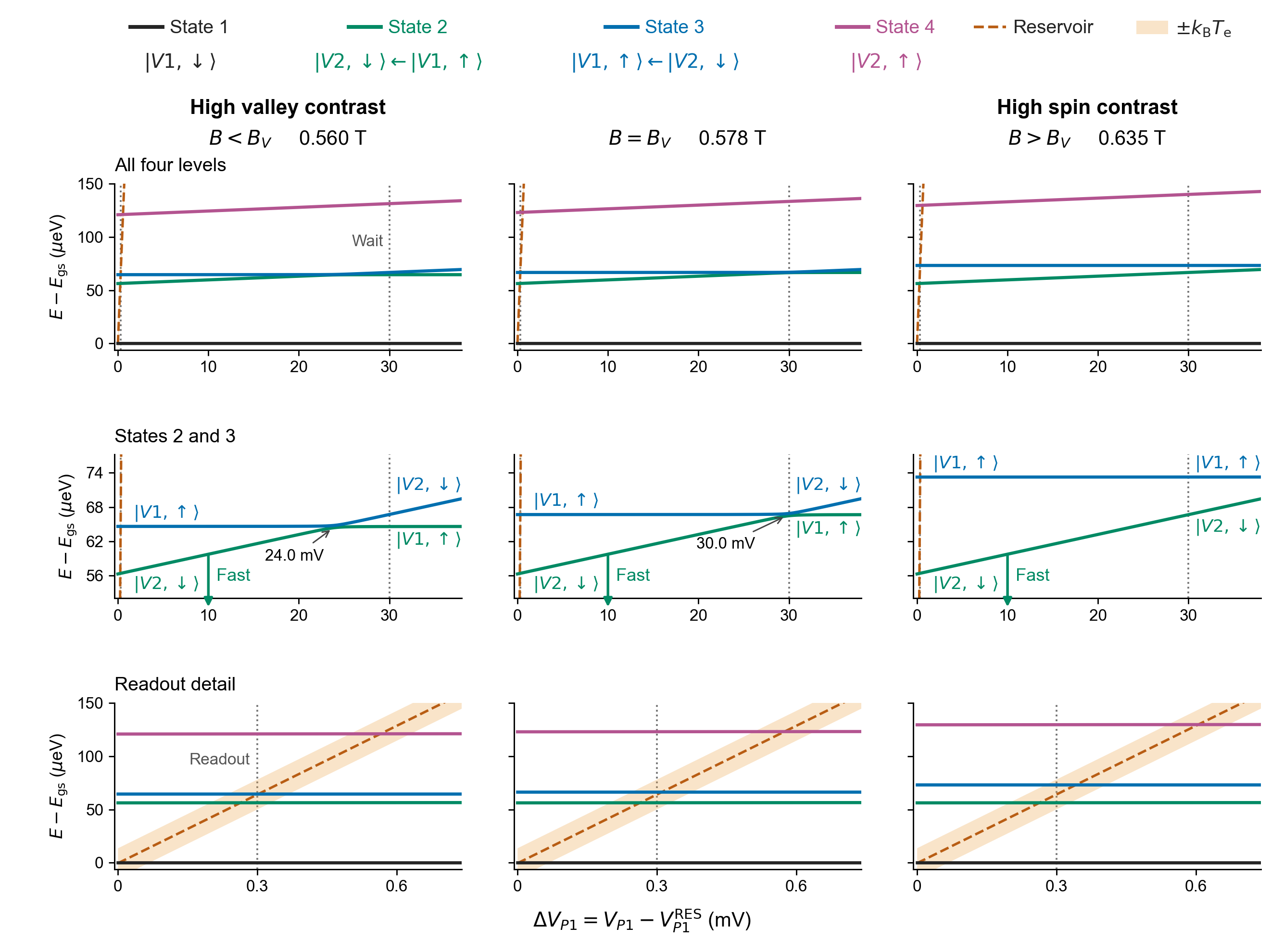}
\caption{Spin--valley energy levels during the Elzerman readout pulse. Calculated spin--valley energies during the P1 pulse at $B=0.560$, $0.578$, and $0.635\,\mathrm{T}$ (left to right). $B_V=0.578\,\mathrm{T}$ is the anticrossing field at the wait point. The plunger-gate voltage offset is $\Delta V_{P1}=V_{P1}-V_{P1}^{\mathrm{RES}}$, where $V_{P1}^{\mathrm{RES}}$ is the resonant tunneling voltage.
(a) The top row shows all four states, numbered in order of increasing energy and shown in black, green, blue, and magenta, respectively. The solid curves show their energies relative to the ground-state energy $E_{\mathrm{gs}}$ at each P1 voltage. The orange dashed line shows the reservoir chemical potential on the same scale. Its height at zero P1 offset is set by the 50:50 charge balance at $V_{P1}^{\mathrm{RES}}$. The shaded band spans $\pm k_BT_e$ ($\pm14\,\mu\mathrm{eV}$) about the dashed line, with $k_B$ the Boltzmann constant and $T_e$ the electron temperature. The gray dotted lines mark readout at $+0.300\,\mathrm{mV}$ and wait at $+30\,\mathrm{mV}$ relative to $V_{P1}^{\mathrm{RES}}$.
(b) The middle row shows states 2 and 3 and their predominant spin--valley character at the ends of the plotted curves. V1 and V2 label the two valleys, and the up and down arrows in the state labels denote spin. The arrows in the legend indicate the mapping for a pulse that crosses the avoided crossing. The gray arrows mark the anticrossings at $24.0$ and $30.0\,\mathrm{mV}$, and the green downward arrows indicate fast valley relaxation. The pulse from wait to readout passes through the avoided crossing at $0.560\,\mathrm{T}$, starts at its center at $0.578\,\mathrm{T}$, and stays on one side at $0.635\,\mathrm{T}$.
(c) The bottom row shows an expanded view near readout. At the readout voltage, states 2 and 3 lie within the thermal band, state 4 lies above it, and state 1 lies below it.}
\label{fig:readout_p1}
\end{figure}

\section{Model of rabi enhancement}

We model the system in the spin–valley basis $\{\ket{V1,\downarrow}, \ket{V1,\uparrow}, \ket{V2,\downarrow}, \ket{V2,\uparrow}\}.$

The static Hamiltonian including the Zeeman splitting and spin–valley-induced hybridization is written as \text{\citesupp{si-EDSRLeti,si-Huang_PhysRevB.90.235315}}
\begin{equation}
H_0+H_{\mathrm{VSO}}=
\begin{bmatrix}
E_{V1}-\tfrac{1}{2}g\mu_B B & 0 & 0 & -\lambda' \\
0 & E_{V1}+\tfrac{1}{2}g\mu_B B & \lambda^{*} & 0 \\
0 & \lambda & E_{V2}-\tfrac{1}{2}g\mu_B B & 0 \\
-\lambda'^{*} & 0 & 0 & E_{V2}+\tfrac{1}{2}g\mu_B B
\end{bmatrix},
\end{equation}
where $E_{V1}$ and $E_{V2}$ are the valley energies, $g$ is the Landé $g$-factor, $\mu_B$ is the Bohr magneton, $B$ is the external magnetic field, and $\lambda$ is the spin–orbit coupling matrix element between $\ket{V1,\uparrow}$ and $\ket{V2,\downarrow}$. This gives rise to four energy-levels, ordered by their energies $(E_1,E_2,E_3,E_4)$. The energy-level diagram is depicted in Fig.~\ref{fig:1}(d), and the levels 2 and 3 are admixed by $\lambda$.

The microwave antenna generates an oscillating transverse magnetic field, which drives a spin-flip transition
\begin{equation}
H_B(t) = \frac{1}{2}g\mu_B B_{\perp}^{ac} \cos(\omega t+\phi_B)\,\hat{B},
\end{equation}
with
\begin{equation}
\hat{B}=
\begin{bmatrix}
0 & 1 & 0 & 0 \\
1 & 0 & 0 & 0 \\
0 & 0 & 0 & 1 \\
0 & 0 & 1 & 0
\end{bmatrix},
\end{equation}
where $B_{\perp}^{ac}$ is the oscillating magnetic field amplitude perpendicular to the static field and $\phi_B$ is the phase of the magnetic drive.

In addition to an a.c. magnetc field, the antenna will also generate an oscillating electric field, which can drive an electric-dipole transition. This field may arise from capacitive coupling between the antenna and nearby gates \citesupp{si-Dehollain2013}. We write the electric drive as
\begin{equation}
H_E(t) = e\,\delta V_G \cos(\omega t+\phi_E)\,\hat{D},
\end{equation}
where $\delta V_G$ is the oscillating gate voltage, $\phi_E$ is the phase of the electric drive and $\hat{D}$ is the dipole operator
\begin{equation}
\hat{D}=
\begin{bmatrix}
D_{V1,V1} & 0 & D_{V1,V2} & 0 \\
0 & D_{V1,V1} & 0 & D_{V1,V2}^* \\
D_{V1,V2}^* & 0 & D_{V2,V2} & 0 \\
0 & D_{V1,V2} & 0 & D_{V2,V2}
\end{bmatrix}.
\end{equation}
Here $D_{Vi,Vi}$ are dimensionless intra-valley matrix elements of the gate drive (nonzero for asymmetric confinement) and $D_{V1,V2}$ is the corresponding inter-valley matrix element.\citesupp{si-PhysRevB.97.155433}

The a.c. magnetic and electric drives can acquire a relative phase at the qubit due to different microwave transfer functions and coupling pathways.  In addition, the transition matrix element of $\hat{D}$ can carry an intrinsic phase set by the device-dependent valley-orbit/spin-orbit physics and the chosen phase convention. For a given transition, these contributions enter only through an effective relative phase, $\Delta\phi$ between the effective electric-dipole and magnetic matrix elements.

Within the rotating-wave approximation, assuming initialization into $\ket{1}$, the Rabi frequencies are
\begin{align}
    hf_{12} &= \left|\,e\delta V_G\,\bra{2}\hat{D}\ket{1}+\exp{(i\Delta\phi)}\,\frac{1}{2}g\mu_B B^{ac}_\perp\,\bra{2}\hat{B}\ket{1}\,\right| ,\\
    hf_{13} &= \left|\,e\delta V_G\,\bra{3}\hat{D}\ket{1}+\exp{(i\Delta\phi)}\,\frac{1}{2}g\mu_B B^{ac}_\perp\,\bra{3}\hat{B}\ket{1}\,\right|.
\end{align}

Let $hf_{E,1n}\equiv e\delta V_G\,\bra{n}\hat{D}\ket{1}$ for $n\in\{2,3\}$ and $hf_{B,1n}\equiv\frac{1}{2}g\mu_B B^{\mathrm{eff}}_\perp\,\bra{n}\hat{B}\ket{1}$, absorbing the relative phase into the effective transverse field \(B_\perp^{\mathrm{eff}}\), and write the inter-valley electric-drive prefactor as $e\,\delta V_G\,D_{V1,V2}^*=\mu_D E_{\mathrm{ac}}$, where $\mu_D=e\bra{V_2}r_D\ket{V_1}$ is the inter-valley electric-dipole matrix element and $r_D$ is the electron position coordinate along the local electric field. The Rabi frequency is then
\begin{align}
hf_{1n}&\approx\left|\,hf_{E,1n}+hf_{B,1n}\,\right|\,=\left|\,e\delta V_G\,\bra{n}\hat{D}\ket{1}+\frac{1}{2}g\mu_B B^{\mathrm{eff}}_\perp\,\bra{n}\hat{B}\ket{1}\,\right|\\
&=\left|\,\,\mu_D E_{\mathrm{ac}}\,\bra{n}(\ket{V1}\bra{V2} + \ket{V2}\bra{V1})\ket{1}\,+\,\frac{1}{2}g\mu_B B^{\mathrm{eff}}_{\perp}\,\bra{n}(\ket{\uparrow}\bra{\downarrow}+\ket{\downarrow}\bra{\uparrow})\ket{1}\,\right|
\end{align}

The ground-state $\ket{1} \approx \ket{V1,\downarrow}$. The upper energy levels $\ket{2},\ket{3}$ are admixtures of $\ket{V1,\uparrow},\ket{V2,\downarrow}$. Below the anti-crossing, $B<B_V$, $f_{12}$ is measured and above the anti-crossing $f_{13}$ is measured.

\section{Angle dependence of inter-valley spin coupling \texorpdfstring{$\lambda$}{lambda}}
\label{sec:lambda}

We can write the Hamiltonian for interface spin-orbit interaction in the coordinate system with $x\parallel[110]$, $y\parallel[\bar{1}10]$, and $z\parallel[001]$ as\citesupp{si-GuoPhysRevLett.124.257701}: %an effective B-field $\vec{b}$
\begin{equation}
    H_{VSO}=(\alpha_{VSO}-\beta_{VSO})k_{y}\,\sigma_{x}-(\alpha_{VSO}+\beta_{VSO})k_{x}\,\sigma_{y}
    \equiv \mu_B\vec{b}_{VSO}\cdot\boldsymbol{\sigma},
\end{equation}
where $\alpha_{VSO},\beta_{VSO}$ are the Rashba, and Dresselhaus coefficients. The canonical momentum $p=k -eA$  has a wave-vector, and magnetic field term. The k-term leads to an effective in-plane spin-orbit field $\vec{b}_{VSO}=(b_x,b_y,0)$ with $b_x=(\alpha_{VSO}-\beta_{VSO})k_{y}$, $b_y=-(\alpha_{VSO}+\beta_{VSO})k_{x}$, $b_z=0$, and the A-term a shift in the g-factor.

The coupling strength $\lambda$ is the component of $\vec{b}$ that is transverse to the direction of the external magnetic field $\hat{n}(\theta,\phi)$:
\begin{equation}
\lambda^{2}(\theta,\phi)=\mu_B^2\left|\vec{b}\times\hat{n}(\theta,\phi)\right|^{2}
=\mu_B^2\left|\hat{n}_{z}\right|^{2}\left(\left|b_{x}\right|^{2}+\left|b_{y}\right|^{2}\right)
+\mu_B^2\left|b_{x}\hat{n}_{y}-b_{y}\hat{n}_{x}\right|^{2}.
\end{equation}

If $\hat{n}(\theta,\phi)=\big(\cos\theta\cos\phi,\ \cos\theta\sin\phi,\ \sin\theta\big)$, $\vec{b}=\big(b\cos\chi\,\exp{(i\xi)},\ b\sin\chi\,\exp{(-i\xi)},\ 0\big)$ then:
\begin{eqnarray}
\lambda^2(\theta,\phi) = (\mu_Bb)^2\left( \sin^2(\theta) + \frac{\cos^2(\theta)}{2}\left(  1- A_{\lambda}\cos(2\phi - \phi_0)\right)\right), \\
A^2_{\lambda} = \cos^2(2\chi) + \cos^2(2\xi)\sin^2(2\chi), \\
\tan(\phi_0) = \tan(2\chi)\cos(2\xi).
\end{eqnarray}

For $\phi_0=-47^{\circ}$, $A_{\lambda}=0.91$, $\chi=25.9^{\circ}$, and $\xi=\pm 74.3^{\circ}$.

\section{Model of g-factor anisotropy}
\subsection{g-tensor definition}
For an electron in a magnetic field, the Hamiltonian can be expressed as
\begin{equation}
H_B=\frac{1}{2}\mu_B \boldsymbol{\sigma}^{T}\hat{g}\vec{B} \equiv \frac{1}{2}\mu_B \boldsymbol{\sigma}^{T}\vec{b} 
\end{equation}
where $\boldsymbol{\sigma}^{T}$ is a row vector $[\sigma_x,\sigma_y,\sigma_z]$, $\vec{B}$ is a column vector $[B_x,B_y,B_z]$, and $\hat{g}$ is a $3\times3$ g-factor tensor matrix, and the Zeeman-splitting $\Delta E = \mu_B\vert \hat{g}\vec{B}\vert$.

The angular dependence of the Zeeman splitting can be expressed as:
\begin{equation}
\Delta E^2=\mu_B^2\left(\vec{B}^T\hat{g}^T\hat{g}\vec{B}\right)=\mu_B^2\left(\vec{B}^T\hat{G}\vec{B}\right),
\end{equation}
where $\hat{G}=\hat{g}^T\hat{g}$.

If we now write $\vec{B}=B\hat{n}$,
\begin{equation}
    \Delta E=\mu_BBg(\hat{n}),
\end{equation}
where $g(\hat{n})=\frac{|\hat{g}\vec{B}|}{B}=|\hat{g}\hat{n}|=\sqrt{\hat{n}^T\hat{G}\hat{n}}$

\subsection{g-factor in-plane angle dependence}

For ease of comparison with theory \citesupp{si-PhysRevB.98.245424}, we analyze the g-factor in the basis of the principal axes, $(x',y',z)=([100],[010],[001])$.   In the $x'-y'$ plane
\begin{equation}
    \hat{n}(\phi)=(\cos(\phi'),\sin(\phi'),0),
\end{equation}
where $\phi'=\phi +\pi/4$, and $\phi'=0$ is aligned along $[100]$ crystal axis.

\begin{align}
g^2(\hat{n})=&\hat{n}^T\hat{G}\hat{n}\\
=&G_{x'x'}\cos^2(\phi')+G_{y'y'}\sin^2(\phi')+2G_{x'y'}\sin(\phi')\cos(\phi')\\
=&\frac{1}{2}(G_{x'x'}+G_{y'y'})+\sqrt{A_{x'y'}}\cos(2\phi'-\delta_{x'y'}),
\end{align}

\begin{equation}
g(\hat{n}) \approx \sqrt{\frac{1}{2}(G_{x'x'}+G_{y'y'})} + a_{x'y'}\cos{(2\phi'-\delta_{x'y'})},\qquad a_{x'y'}=\frac{1}{2}\sqrt{\frac{A_{x'y'}}{\frac{1}{2}(G_{x'x'}+G_{y'y'})}},
\end{equation}

where
\begin{equation}
A_{x'y'}=\left[\frac{1}{2}(G_{x'x'}-G_{y'y'})\right]^2+G_{x'y'}^2 = \left[\frac{1}{2}(g_{x'x'}^2-g_{y'y'}^2)\right]^2 + (g_{x'x'}g_{x'y'}+g_{y'x'}g_{y'y'})^2
\end{equation}
and
\begin{equation}
\tan{(\delta_{x'y'})}=\left(\frac{G_{x'y'}}{\frac{1}{2}(G_{x'x'}-G_{y'y'})}\right) = 2\frac{g_{x'x'}g_{x'y'}+g_{y'x'}g_{y'y'}}{g_{x'x'}^2-g_{y'y'}^2}
\end{equation}
Assuming $g_{y'z'}=g_{x'z'}=0$, for an in-plane spin-orbit interaction. A value of $\delta_{x'y'}=0,\pi/2 $ corresponds to alignment of the maximum in g-factor with $[100],[110]$ axes, respectively.

A similar expression can be found for a B-field in x-z plane by changing the indices.

\subsection{In-plane anisotropy of difference in g-factor of Valley states.}
\label{sec:Delta_g}
The difference in g-factors of the valley states is
\begin{equation}
\Delta g_V(x'y')=g_{V1}-g_{V2}\approx c+ a_{x'y'}^{V1}\cos{(2\phi'-\delta_{x'y'}^{V1})}-a_{x'y'}^{V2}\cos{(2\phi'-\delta_{x'y'}^{V2})},
\end{equation}
where $c$ is a constant. If we assume the only difference in g-factor tensor is in the off-diagonal terms, and that $g_{x'y'}^{V1}=-g_{x'y'}^{V2}$, since $g_{x'y'}\propto \beta$, where $\beta$ is the Dresselhaus coefficient \citesupp{si-PhysRevB.98.245424}, then $a_{x'y'}^{V1}=a_{x'y'}^{V2}=a_{x'y'}, \delta_{x'y'}^{V1}=-\delta_{x'y'}^{V2}=\delta_{x'y'}$ then
\begin{equation}
\Delta g_V \approx 2a_{x'y'}\sin{(2\phi')}\sin{(\delta_{x'y'})}
\end{equation}
This has maximum amplitude when $\phi'=\pi/4, \phi=0$ along crystal axes $[110]$ or $[\bar{1}10]$, as observed.
\subsection{Anisotropy of in-plane mean g-factor}
\label{sec:mean_g}

The mean g-factor of the valley states is
\begin{equation}
    \bar{g}_V(x'y')=\frac{g_{V1}+g_{V2}}{2}\approx c+ \frac{1}{2}\left[a_{x'y'}^{V1}\cos{(2\phi'-\delta_{x'y'}^{V1})}+a_{x'y'}^{V2}\cos{(2\phi'-\delta_{x'y'}^{V2})}\right],
\end{equation}
where $c$ is a constant. As before, if we assume only difference in g-factor tensor is in the off-diagonal terms, and that $g_{x'y'}^{V1}=-g_{x'y'}^{V2}$, since $g_{x'y'}\propto \beta$, where $\beta$ is the Dresselhaus  \citesupp{si-PhysRevB.98.245424}, then $a_{x'y'}^{V1}=a_{x'y'}^{V2}=a_{x'y'}, \delta_{x'y'}^{V1}=-\delta_{x'y'}^{V2}=\delta_{x'y'}$ then
\begin{equation}
\bar{g}_V \approx c + a_{x'y'}\cos{(2\phi')}\cos{(\delta_{x'y'})}
\end{equation}
This has maximum amplitude when $\phi'= 0,\phi =-\pi/4$ along crystal axes $[100]$ or $[010]$, as observed.

\section{Additional Figures}

\begin{figure}[h!]
\centering
\includegraphics[width=\linewidth/2]{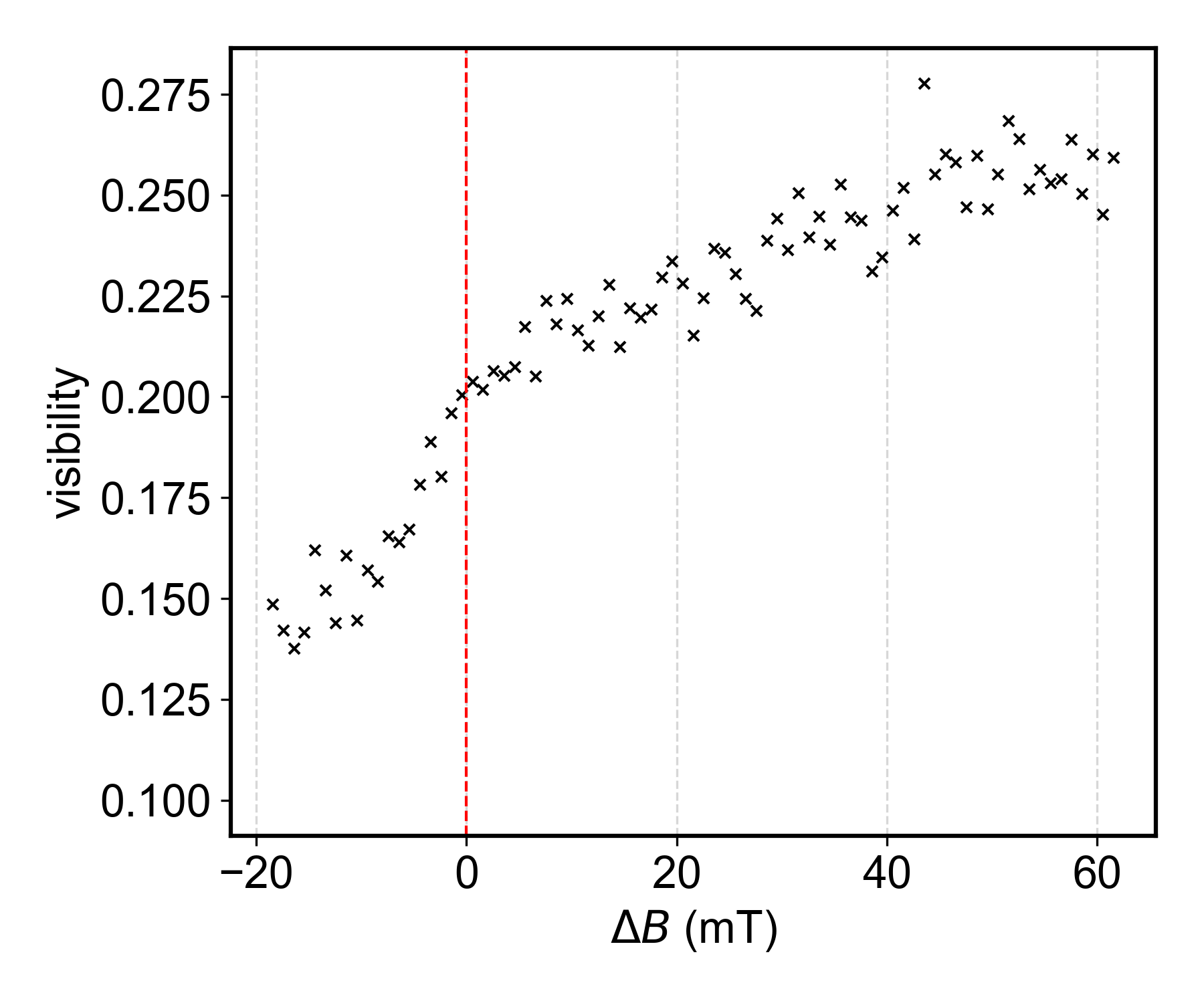}
\caption{Blip probability during the relaxation measurement. The black crosses show the fitted blip probability at zero wait time. The red dashed line marks zero detuning, \(\Delta B=B-B_V=0\), where \(B_V\) is the anticrossing field. In the high-field limit, the blip probability is around \(25\%\). Relaxation during the pulse to the load point and the readout point can reduce the measured blip probability even at nominally zero wait time. At lower magnetic fields, the blip probability drops further.}
\label{fig:vis}
\end{figure}

Supplementary Fig.~\ref{fig:vis} shows the fitted blip probability at zero wait time.

\begin{figure}[h!]
\centering
\includegraphics[width=\linewidth]{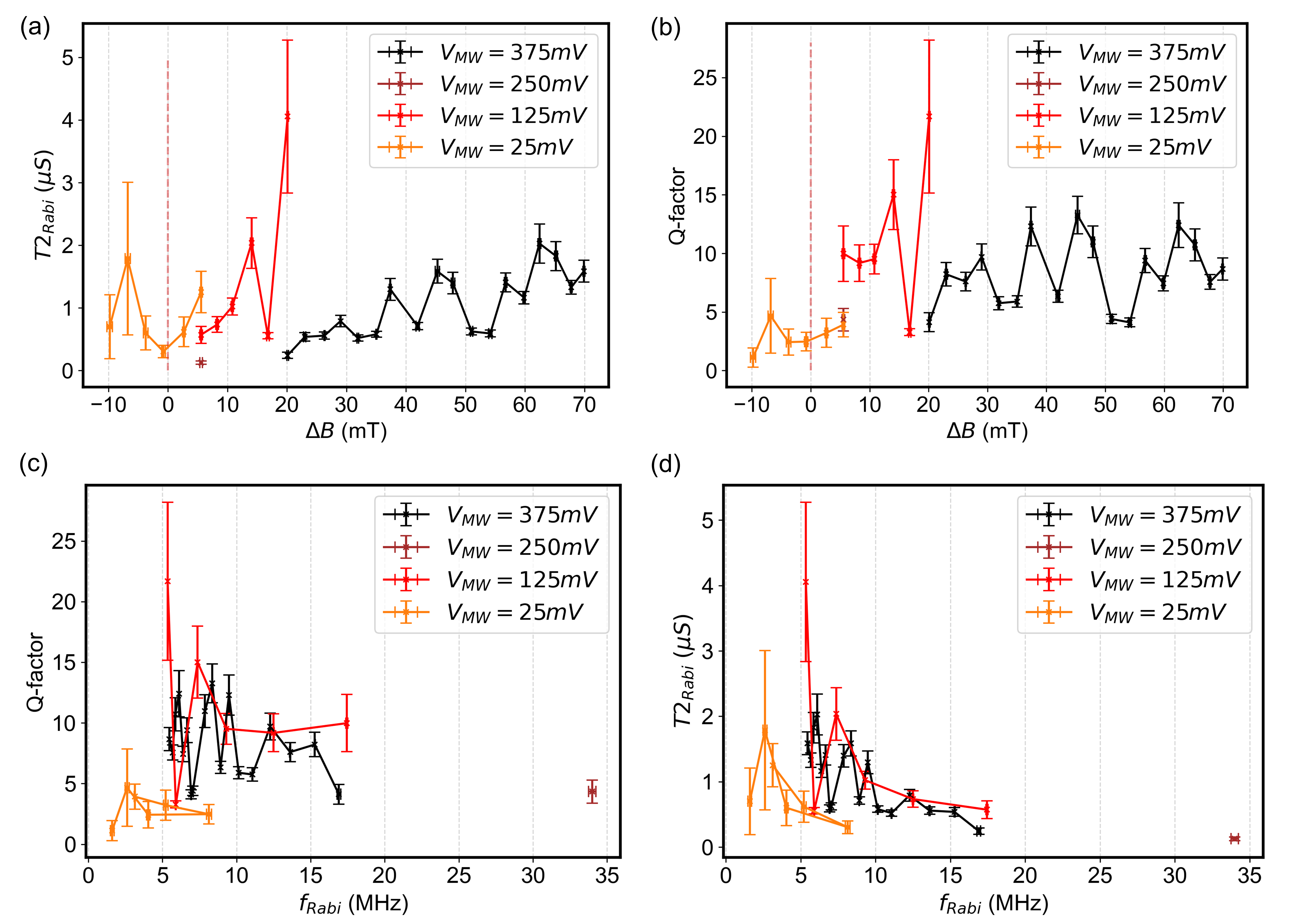}
\caption{Rabi damping times and quality factors near the spin--valley anticrossing. (a) Rabi quality factor $Q=f_{\mathrm{Rabi}}T_{2,\mathrm{Rabi}}$ versus detuning $\Delta B=0.635\,\mathrm{T}-B_V$, where $B_V$ is the anticrossing field. (b) Rabi damping time $T_{2,\mathrm{Rabi}}$ versus detuning. (c) Q-factor versus Rabi frequency $f_{\mathrm{Rabi}}$. (d) Rabi damping time versus Rabi frequency. Black, brown, red, and orange crosses correspond to microwave source voltages $V_{\mathrm{MW}}$ of $375$, $250$, $125$, and $25\,\mathrm{mV}$, respectively. The red dashed lines in (a) and (b) mark zero detuning. Error bars show uncertainties from the Rabi fits and the valley-splitting calibration. The damping time fluctuates a lot due to slow noise in the device, and a larger source voltage degrades the damping time \protect\citesupp{si-peetroons2025highfidelityqubitcontrol}. The data suggest an optimum Q-factor at $\Delta B\approx20\,\mathrm{mT}$. The damping time suggests an optimum value, a trade-off between greater dynamical decoupling from noise in the qubit frequency at high Rabi frequencies, and greater susceptibility of the Rabi frequency to charge-noise-induced fluctuations in the valley splitting.}
\label{fig:Q-factor}
\end{figure}

\newpage
Supplementary Fig.~\ref{fig:Q-factor} here shows the Q-factor and $T2_{Rabi}$ dependence on the detuning and the Rabi frequency in the Rabi measurement in Fig.~\ref{fig:rabi}(c).

\begin{figure}[h!]
\centering
\includegraphics[width=\linewidth]{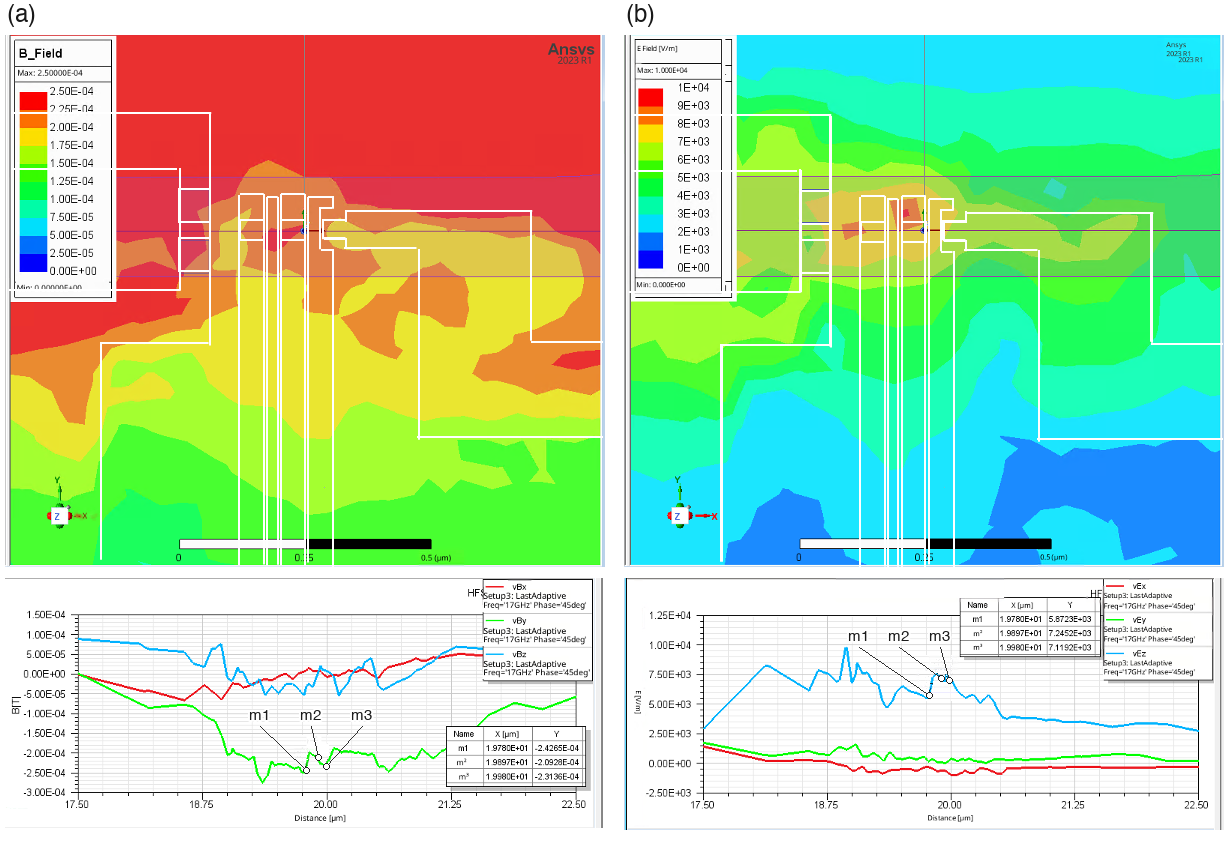}
\caption{Simulated microwave fields at 17 GHz. (a) Magnetic field map (top) and line scans of \(B_x\), \(B_y\), and \(B_z\) (bottom). (b) Electric field map (top) and line scans of \(E_x\), \(E_y\), and \(E_z\) (bottom). The map colors run from blue to red with increasing field strength, in tesla in (a) and volts per metre in (b). In both line scans, the red, green, and blue traces show the \(x\), \(y\), and \(z\) components, respectively. The line scans were evaluated at a simulation phase of \(45^\circ\) and a drive power of \(0.1\,\mathrm{mW}\). Markers \(m_1\), \(m_2\), and \(m_3\) identify the locations of the single-electron transistor (SET) sensor and plunger gates P1 and P2, respectively. At P1, the extracted values are \(B_y=-209.28\,\mu\mathrm{T}\), \(B_z\approx-5\,\mu\mathrm{T}\), and \(E_z=7.2452\,\mathrm{kV/m}\), where \(B_y\) and \(B_z\) are the transverse magnetic field components and \(E_z\) is the vertical electric field. With \(|B_\perp|=\sqrt{B_y^2+B_z^2}\), these values give \(|E_z|/|B_\perp|\simeq34.6\,\mathrm{kV\,m^{-1}\,mT^{-1}}\).}
\label{fig:hfss_17ghz}
\end{figure}

\newpage
Supplementary Fig.~\ref{fig:hfss_17ghz} here shows the electromagnetic simulation of the antenna and gate layout at $17~\mathrm{GHz}$.

\clearpage
\bibliographystylesupp{naturemag-doi}
\bibliographysupp{supplementary}

\end{document}